\documentclass{article}
\usepackage[utf8]{inputenc}
\usepackage[margin=1in]{geometry}

\usepackage{url}
\usepackage{lineno}
\usepackage{overpic}
\usepackage{algorithm}
\usepackage{algorithmic}
\usepackage{amsmath}
\usepackage{amssymb}
\usepackage{mathrsfs}
\usepackage{soul}
\usepackage{authblk}
\usepackage{color}
\usepackage{ulem}
\usepackage[round]{natbib}

\title{\vspace{-1.5cm}A mushy source for the geysers of Enceladus}
\author[1]{Colin R. Meyer\thanks{colin.r.meyer@dartmouth.edu}}
\author[1]{Jacob J. Buffo}
\author[2]{Francis Nimmo}
\author[3]{Andrew J. Wells}
\author[4]{Samuel Boury}
\author[1]{Tara C. Tomlinson}
\author[3]{Jamie R. G. Parkinson}
\author[5]{Geoffrey M. Vasil}
\affil[1]{Thayer School of Engineering, Dartmouth College, Hanover, NH 03755}
\affil[2]{Department of Earth \& Planetary Sciences, University of California, Santa Cruz, CA 95064 USA}
\affil[3]{Atmospheric, Oceanic \& Planetary Physics, Department of Physics, University of Oxford, OX1 3PU, UK}
\affil[4]{Courant Institute of Mathematical Sciences, New York University, New York, NY 10012}
\affil[5]{School of Mathematics, University of Edinburgh, EH9 3FD, UK}

\begin{document}

\maketitle

\begin{abstract}
Enceladus is a primary target for astrobiology due to the $\rm H_2O$ plume ejecta measured by the Cassini spacecraft and the inferred subsurface ocean sustained by tidal heating. Sourcing the plumes via a direct connection from the ocean to the surface requires a fracture through the entire ice shell ($\sim$10~km). Here we explore an alternative mechanism in which shear heating within shallower tiger stripe fractures produces partial melting in the ice shell and interstitial convection allows fluid to be ejected as geysers. We use an idealized two-dimensional multiphase reactive transport model to simulate the thermomechanics of a mushy region generated by an upper bound estimate for the localized shear heating rate in a salty ice shell. From our simulations, we predict the temperature, porosity, salt content, melting rate, and liquid volume of an intrashell mushy zone surrounding a fracture. We find that the rate of internal melting can match the observed $\rm H_2O$ eruption rate and that there is sufficient brine volume within the mushy zone to sustain the geysers for $\sim350$~kyr without additional melting. The composition of the liquid brine is, however, distinct from that of the ocean, due to partial melting. This shear heating mechanism for geyser formation applies to Enceladus and other icy moons and has implications for our understanding of the geophysical processes and astrobiological potential of icy satellites. 
\end{abstract}
\vspace{1cm}


After the discovery of eruptive jets on Saturn's moon Enceladus \citep{Por2006}, two distinct mechanisms were proposed to explain their origin. First, \citet{Nim2007} proposed that tidally driven shear heating along the south pole fractures causes solid ice to sublimate, generating a vapor source for the plumes. Second, \citet{Hur2007} argued that tidally driven crack opening exposes a subsurface ocean directly to space, later reinforced by \citet{Kit2016}. It is the second explanation that is now generally favored for three reasons: first, the librations of Enceladus are consistent with the existence of a global liquid ocean \citep{Tho2016}; second, the plume is modulated on an orbital period, as originally predicted by \citet{Hur2007} and later corroborated by several independent groups \citep{Hed2013,Nim2014,Ing2017}; and third, the erupted material is salty \citep{Pos2009,Pos2011}. This last observation is the strongest indication of a connection between the subsurface ocean and the eruptive material, as most models assume a salt-free ice shell. \citet{Kit2016} showed that turbulent dissipation of tidally driven water within a fracture is sufficient to maintain an open crack from the ocean to the surface. On the other hand, \citet{Nak2016} found that refreezing condensation onto the fracture walls above the water level produces thin fractures ($\sim$0.1~m wide) that close on timescales of less than a year, making it difficult to explain how near-surface fractures stay open. Motivated by Cassini's Composite Infrared Spectrometer (CIRS) observations, \citet{Abr2009} constrained the width of the fracture zone in a two-dimensional thermal model and found that a narrow crack with a temperature that is far below the melting point of pure ice (i.e. $\sim225$ K crack temperature for their geometry) fits the observed thermal spectrum. Considering the difficulties of fracturing through a $\sim$10~km ice shell (see supplement) and rapid closure, it is of interest to investigate whether we can find a mechanism explaining the salty jets which does not require direct connection to a subsurface ocean. In this paper, we demonstrate that shear heating along a tiger stripe fracture in a salty ice shell can induce localized partial melting and provide an alternative explanation for the observed salty plumes particles that does not rely on a direct connection with the ocean.

In many frozen terrestrial and planetary contexts, strike-slip shear heating along narrow zones of deformation efficiently warms the surrounding ice \citep{Nim2002,Gol2010, Mey2018a}. By modeling the transport of heat and sublimation vapor flow in a pure ice porous region around a tiger stripe fracture, \citet{Nim2007} found a partitioning of sensible and latent heat that explained the vapor transport as well as the thermal anomaly. Observations, however, show that the erupting material consists of pure and salty ice grains as well as molecular hydrogen H$_2$, and silica nanoparticles, providing evidence for an oceanic origin for the plumes and signatures of water-rock interactions \citep{Pos2011,Hsu2015,Wai2017}. Sublimation cannot produce ice grains with dissolved salts or dust. A liquid source must therefore be responsible for the formation of the geyser ice grains, i.e. either a direct connection to the ocean or a reservoir within the shell. Freezing of the shell from a parent ocean could have trapped salt, suspended dust, and incorporated dissolved gases in the shell \citep{Wai2006,Wai2017,Buf2021b}. Molecular hydrogen H$_2$ could be trapped in near-surface clathrates with stabilization from CH$_4$ and CO$_2$ \citep{Wai2017} or persist as entrained gas bubbles within the ice shell, as found in ice sheets on Earth \citep{Ben1997}. Substantial depletion of the ice shell by diffusion is unlikely: laboratory evidence shows that the diffusion of H$_2$ through ice at Earth temperatures (with diffusivity $D_{H_2}\sim 10^{-11}$ m$^2$ s$^{-1}$) is at least an order of magnitude smaller than the first-order estimate of the diffusivity $D \sim h^2/t \sim 7\times10^{-10}$ m$^2$ s$^{-1}$ required for gas to escape from an $h\sim 10$ km shell in $t_{\text{age}}\sim 4.5$ billion years. The diffusivity $D_{H_2}$ is likely even smaller at Enceladus temperatures \citep{Pat2021}, further reducing the possibility of substantial diffusive escape of $\rm H_2$.

\begin{figure}[ht!]
\centering
\begin{overpic}[width=0.5\linewidth]{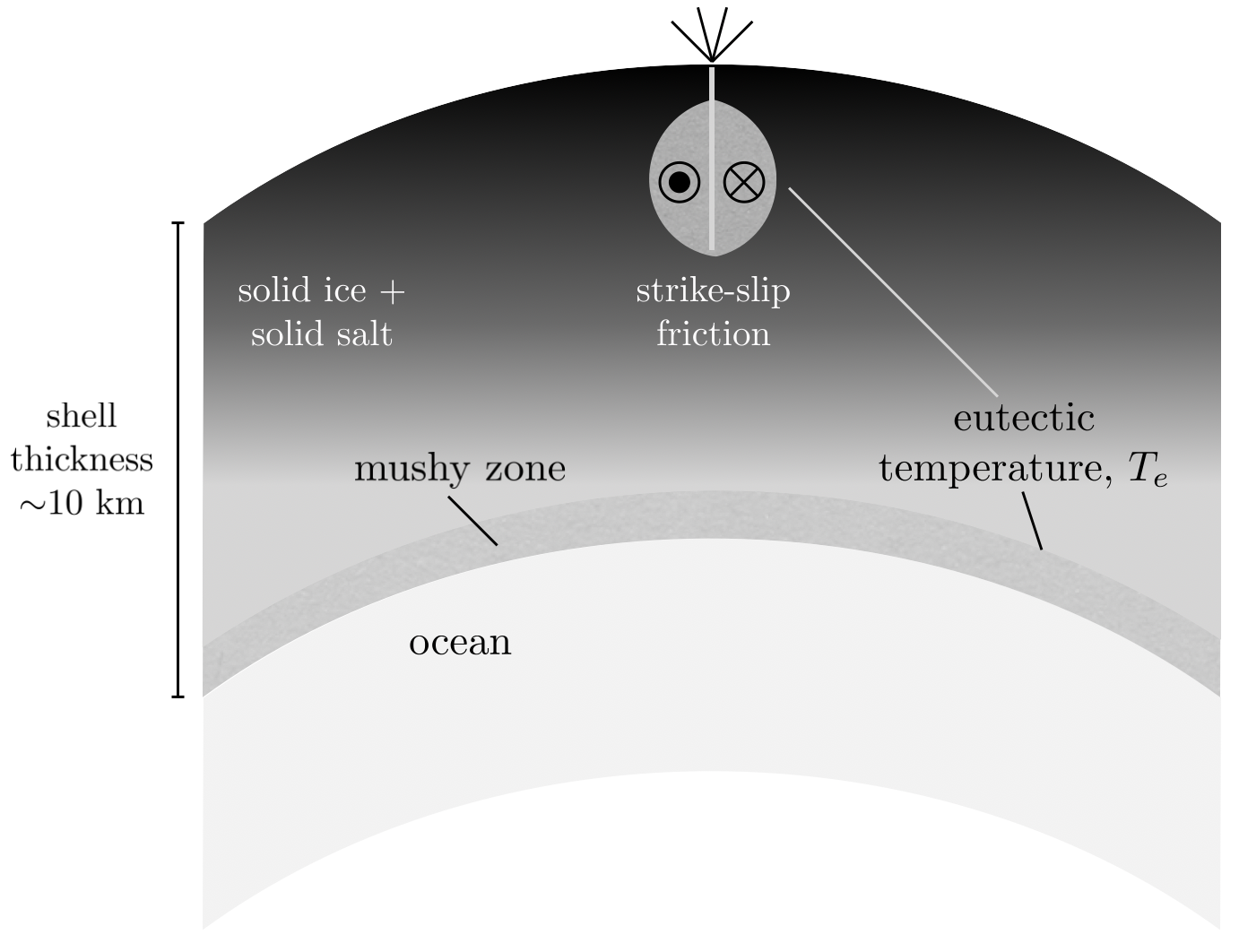}\end{overpic}
\includegraphics[width=0.49\linewidth]{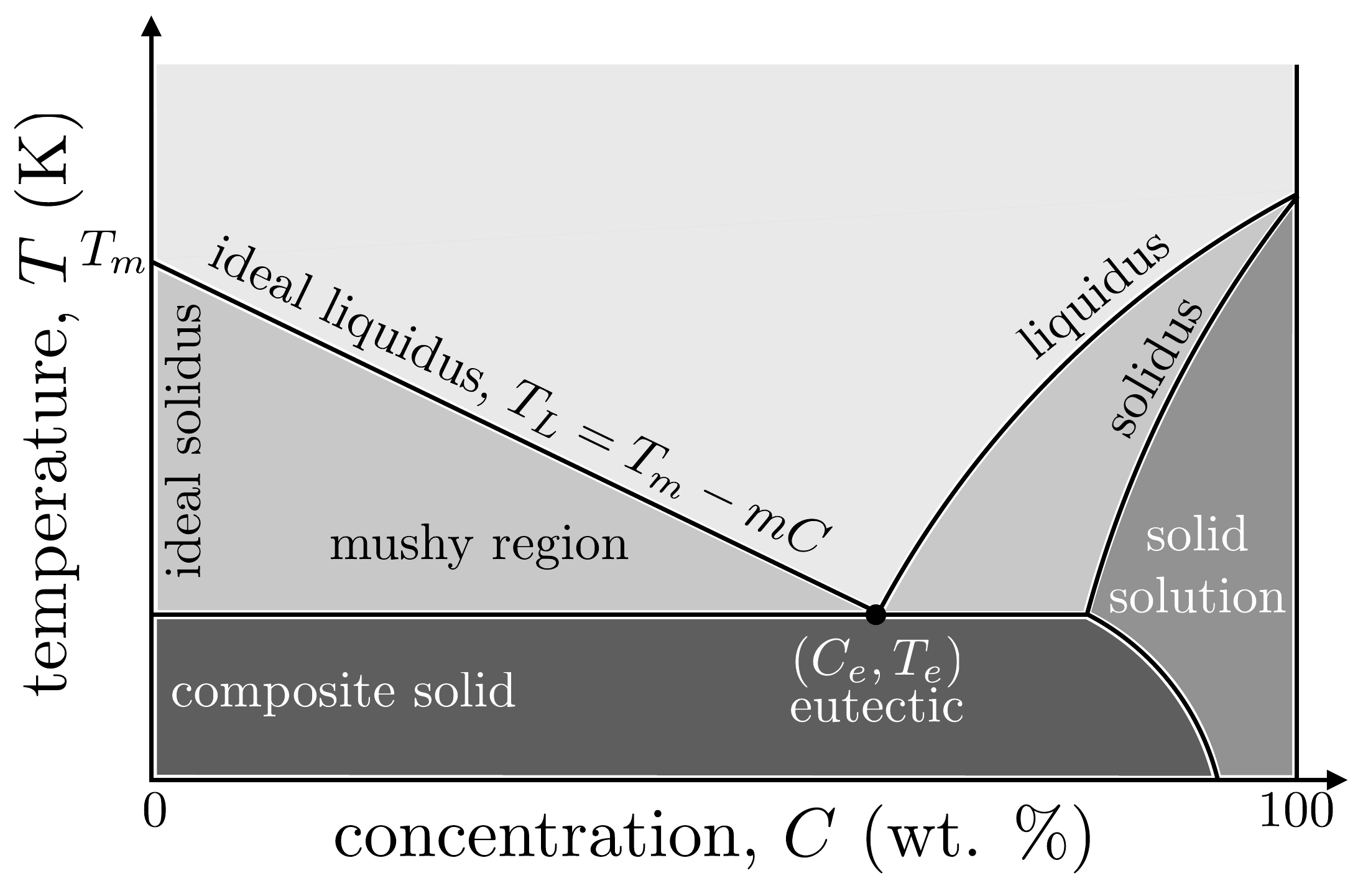}
\caption{Model for a mushy region around the tiger stripes of Enceladus: (left) Schematic for the composition of Enceladus' shell demonstrating strike-slip friction along a tiger stripe fracture. Slip on the finite-length fault varies over time in both directions, although we only depict one direction of the strike-slip motion here. Shear heating on the fracture generates a mushy zone of salty water surrounding ice crystals, providing a liquid source for the geyser material. Darker colors qualitatively correspond to higher expected salt content \citep{Buf2021b}. (right) Idealized equilibrium phase diagram for a eutectic binary alloy (e.g. NaCl and H$_2$O) as a function of temperature and concentration, after \citet{Wor2000}. At temperatures above the liquidus, the mixture is entirely liquid. In the region between the liquidus and solidus, there is two-phase coexistence, i.e. a mushy zone, where saline liquid occupies the interstices of a solid ice matrix. Below the eutectic point $(C_e,T_e)$, a solid forms that is a composite of salt and ice.}
\label{fig:shellschematic}
\end{figure}

Physicochemically heterogeneous materials are common throughout mantles and crusts in planetary geophysics, e.g. magma dynamics in the Earth's mantle \citep{Kat2008a}, sea ice growth \citep{Hun2011,Buf2018}, and marine ice accumulation under ice shelves \citep{Bom1995}. Given the oceanic origin and geophysical reworking of active planetary ice shells (e.g., Europa, Enceladus), it is likely that there is a significant and heterogeneous distribution of salt frozen into the shells of these icy satellites \citep{Pos2011,Tru2019}. Oceanic salt and silica nanoparticles could be incorporated into the ice shell by freezing during Enceladus' formation, or during resurfacing  \citep{Ham2018,Buf2021b,Buf2021a}. Thinner ice shells will have a larger conductive heat flux and faster growth. Since the amount of impurities incorporated increases with the freezing rate, the near-surface ice is likely to be saltier than the ice at greater depths \citep{Buf2021b}. Entrained salt in the shell means that the melting temperature of the ice is depressed: the impurities in heterogeneous systems facilitate partial melting at lower temperatures than the melting point of pure ice. Shear heating along a tiger stripe fracture could therefore induce partial melting within the shell and generate a near-surface salty liquid source for the south polar plume of Enceladus (figure~\ref{fig:shellschematic}).

In a multicomponent system, the eutectic temperature $T_{e}$ is the lowest temperature at which interstitial liquid is stable \citep{Wor2000}. Given that the ocean temperature is above the eutectic, there is likely a thin mushy zone of salty water surrounding ice crystals at the ice-ocean interface of icy satellite shells \citep{Buf2021b}. The minimum depth within the shell where salt exists in a liquid solution is determined by the eutectic temperature, cf. figure \ref{fig:shellschematic} (left). Figure \ref{fig:shellschematic} (right) shows a schematic equilibrium phase diagram for a eutectic binary alloy, such as NaCl and H$_2$O with $T_e = 252$ K. For Enceladus, the rapid loss of heat to outer space leads to near-surface temperatures of about $T_s=75$ K, i.e. below the eutectic, where solid salts can be maintained within the ice shell as a composite phase. At places in the shell where the temperature is above the eutectic (see figure \ref{fig:shellschematic}) the salts are dissolved in the interstitial fluid and occupy the pore space around nearly fresh ice crystals. This region is a mushy zone, a region of two-phase coexistence where the liquid and solid phases are in equilibrium, that corresponds to the region between the solidus and liquidus in figure \ref{fig:shellschematic} (right). Close to the fracture, shear heating can warm the ice above the eutectic temperature $T_e$ (viz. figure \ref{fig:shellschematic}, left) and form a mushy zone. 

\begin{figure}[ht!]
\includegraphics[width=0.44\linewidth]{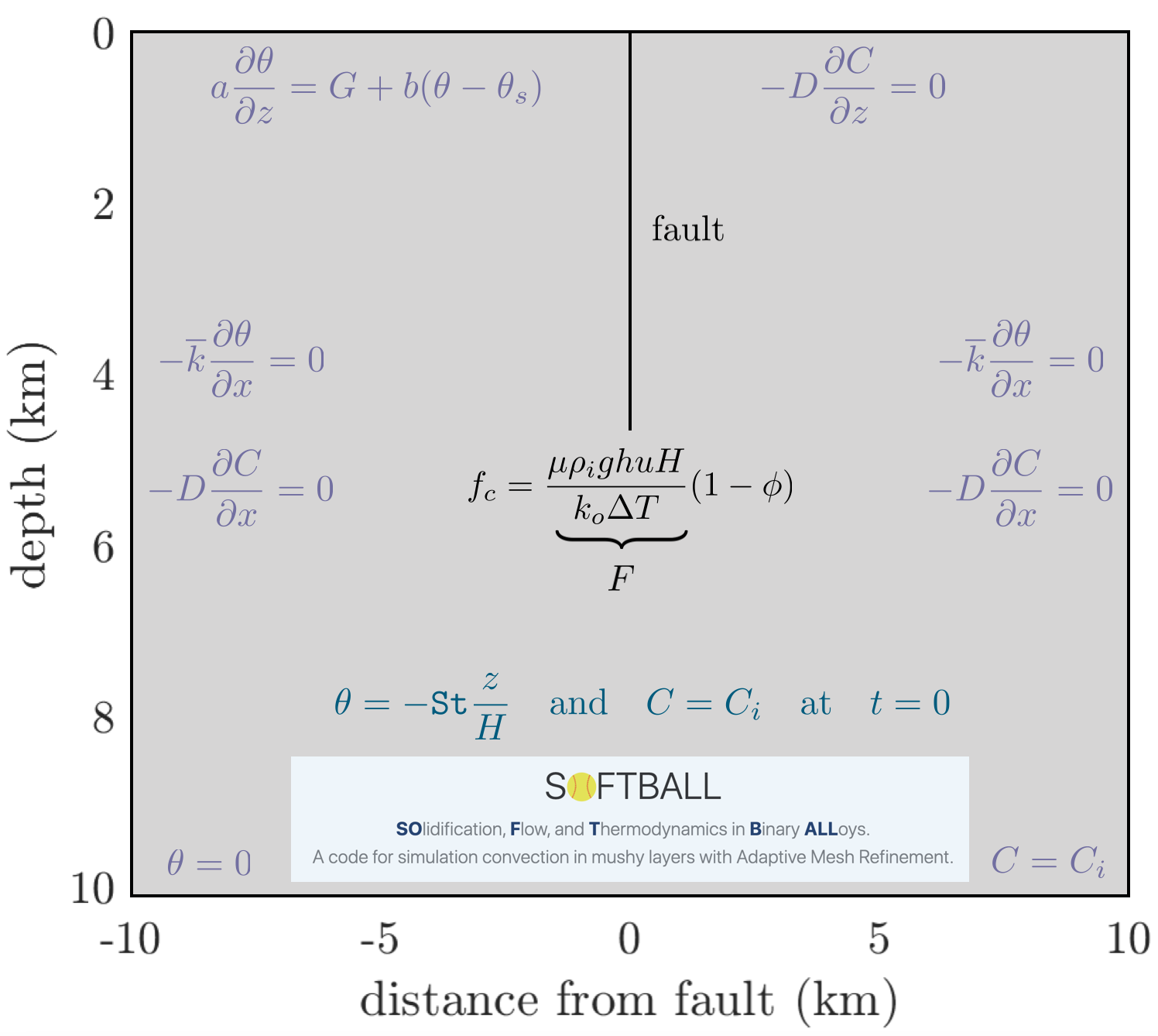}
\includegraphics[width=0.525\linewidth]{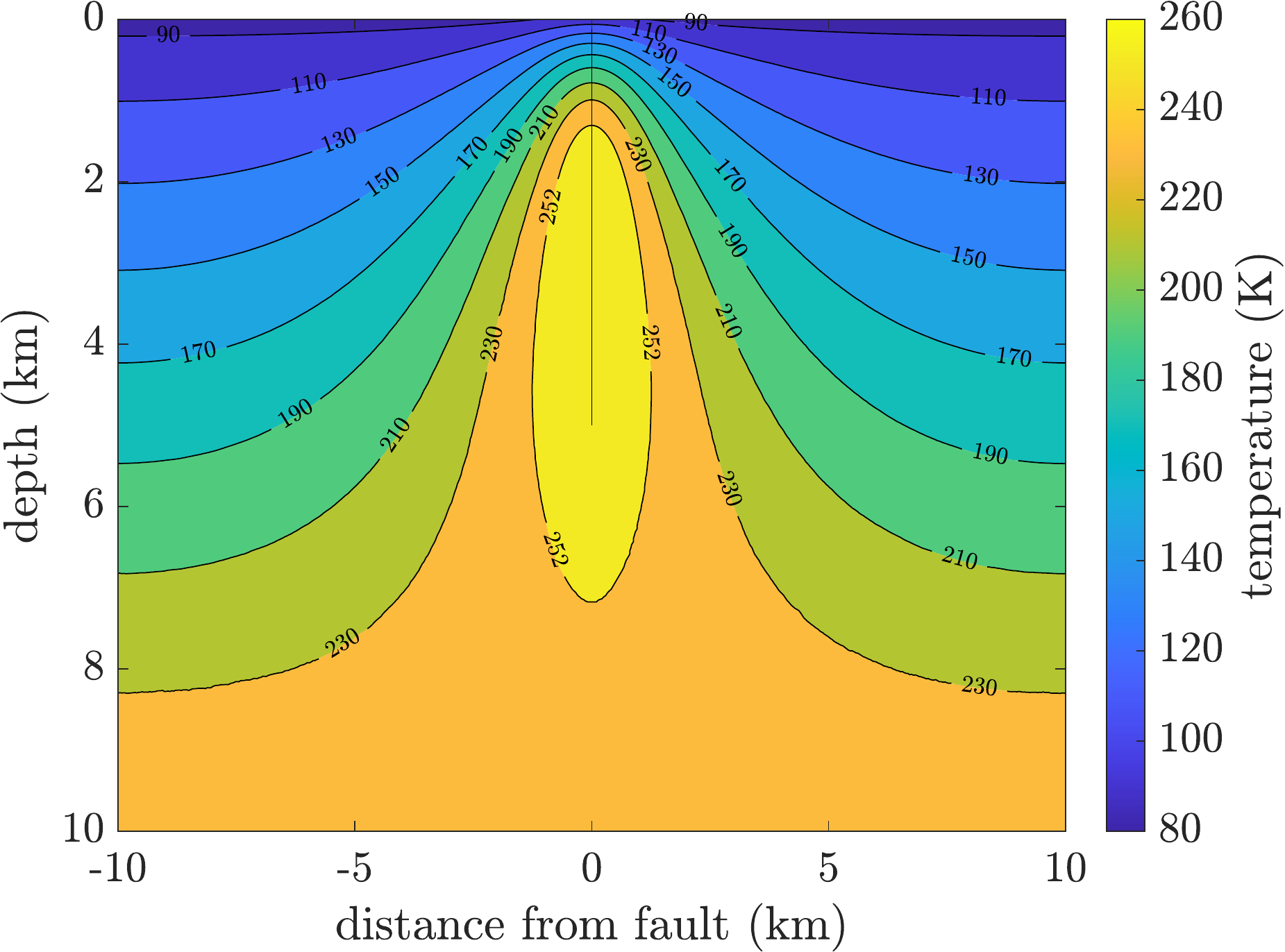}
\caption{Numerical simulations of mushy zones around a fracture. (left) Schematic showing the simulation domain, initial conditions (teal), boundary conditions (purple), and heating along the fracture (black) where $\theta$ is dimensionless temperature, $\phi$ is porosity, $C$ is composition, and $\mathtt{St} = \mathscr{L}/(c_p \Delta T)$ is the Stefan number, with latent heat $\mathscr{L}$ and heat capacity $c_p$. (right) Temperature contours showing the development of a mushy zone ($T \geq T_e=252$ K) around the fracture (dimensionless heating rate, $F=500$, corresponding to a slip rate of 3.4 $\mu$m/s).}
\label{fig:tempcontour}
\end{figure}

We describe the equations for the evolution of a mushy zone in the supplement \citep{Fow1985,Wor2000,Hun2011}. We solve the mushy layer equations numerically in SOFTBALL \citep[SOlidification, Flow, and Thermodynamics in a Binary ALLoy, see methods section;][]{Par2020}. In figure~\ref{fig:tempcontour}, we show the set up of the problem, the key parameters, and the numerically computed temperature field. We consider an isolated fracture in a compositionally homogeneous ice shell (see schematics in figures~\ref{fig:shellschematic} and~\ref{fig:tempcontour}) where the time-averaged effect of many cycles of strike-slip friction (back and forth due to tidal forces) is approximated by steady shear heating localized at the crack. We take a given value of the nondimensional heat flux $F = \mu \rho_i g h u H / (k_o\Delta T)$, with friction coefficient $\mu$, ice density $\rho_i$, gravity $g$, crack depth $h$, slip velocity $u$, ice shell thickness $H$, thermal conductivity $k_o$, and temperature difference $\Delta T$ (see methods and supplement). We then numerically solve for the temperature, salinity, and porosity around the fracture using SOFTBALL (see figure \ref{fig:tempcontour} and methods). We take a nominal slip velocity of 1.8 to 3.4 $\mu$m/s, following \citet{Nim2007} (see supplement for table of parameters). The model conserves mass, salt, energy, and momentum for flow in a reactive porous medium with a porosity dependent permeability. We approximate the densities of ice and water as equal and neglect the viscous deformation of the ice matrix. For the present model, we focus on melt generation and do not include melt extraction along the cracks due to geyser eruptions (see supplement). 

\section*{Results}
We find cold surface temperatures and a mushy zone around the fracture (figure~\ref{fig:tempcontour}, right). The computed temperature profiles are similar to the results of \citet{Kal2016}, \citet{Hamm2020}, and \citet{Ste1996}, which are all for a pure ice shell. In the supplement, we show that the leading-order thermal structure can be understood from steady conduction, with reactive flow causing subsequent modifications to the porosity structure, composition and shape of the inclusion. Using SOFTBALL, we find that there is a significant amount of interstitial brine within the mushy zone. Cold, saline brine sinks in the outer part of the mushy inclusion  driving convection, with warmer and fresher fluid rising near the crack (see vertical velocity arrows in figure \ref{fig:Fvolume} and in supplement figure 4). This is broadly consistent with a simplified model of buoyant convection in a mushy zone by \citet{Bou2021}. The vigor of the convection is characterized by the thermal and compositional Rayleigh numbers \citep[see supplement and][]{Par2020}. Above a critical value, the rising plume becomes unstable. Here we follow \citet{Bou2021} and use a small permeability that results in values of the Rayleigh numbers that are below the threshold for an unstable plume (see supplement). Near the surface, following the model of \citet{Ing2016}, the fluid will boil off and leave the satellite as geysers that may be fresher than the liquid source \citep{Pos2009,Pos2011}. In addition, bubbling formation and bursting near the liquid-space interface explains the presence of salt-poor, organics-rich ice grains \citep{Pos2018}. 

The circulation of brine through the porous ice matrix results in a pooling of salt towards the bottom of the mushy inclusion, with further drainage prevented in our model by the colder impermeable ice below. This region thus has enhanced phase-weighted salinity, and more easily melts, eventually leading to a region of nearly pure fluid near the bottom of the crack (see figure \ref{fig:Fvolume}, right). Accounting for deformation of the ice matrix, it is possible that the pure fluid region could propagate downward as a dike, thereby initiating a connection from the surface to the ocean. In the supplement, we quantify the timescale for dike propagation based on \citet{Lis1990} and find that it is comparable to the timescale of our simulations. At the same time, plume eruptions would reduce the volume of melt in the mushy zone, reducing the likelihood of dike formation. At present, however, our model neglects the potential for sinking of the melt-rich pocket via fracturing or viscous deformation of the surrounding ice, due to the density difference between water and ice \citep[cf.][]{Kal2016}.

We quantify the amount of interstitial water in the mushy zone as the total brine volume  
\begin{equation}
v = \ell \int_{A}\phi~dS, 
\end{equation}
where $\ell = 500$ km is the length of the tiger stripe fractures perpendicular to the model domain, and $A$ the area of the domain (see figure~\ref{fig:Fvolume}). Here we choose $\ell$ to be the combined length of all four tiger stripe fractures, so that our calculations represent the total effects rather than the dynamics of a single fracture.

\begin{figure}[ht!]
\includegraphics[width=0.515\linewidth]{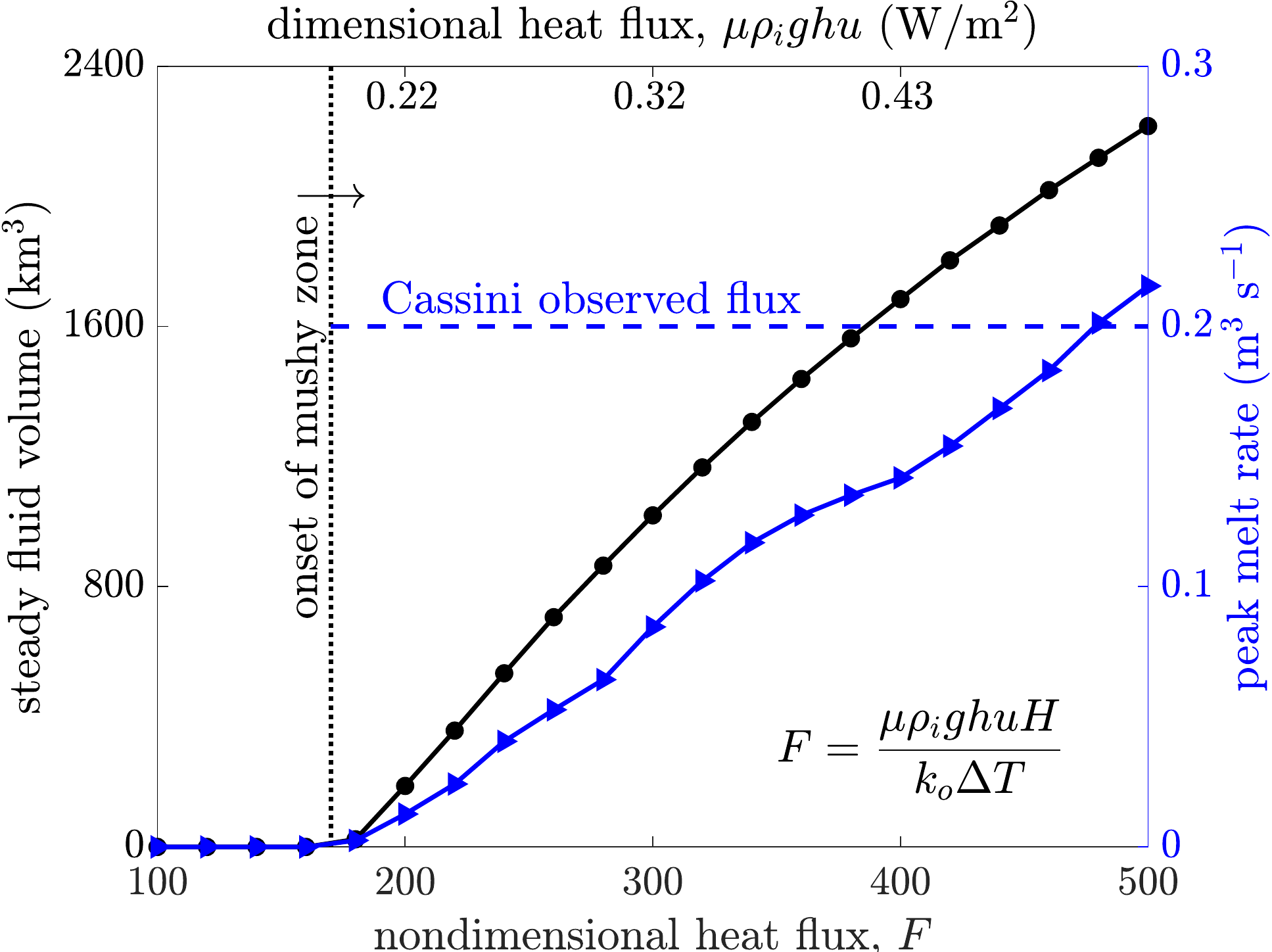}
\includegraphics[width=0.485\linewidth]{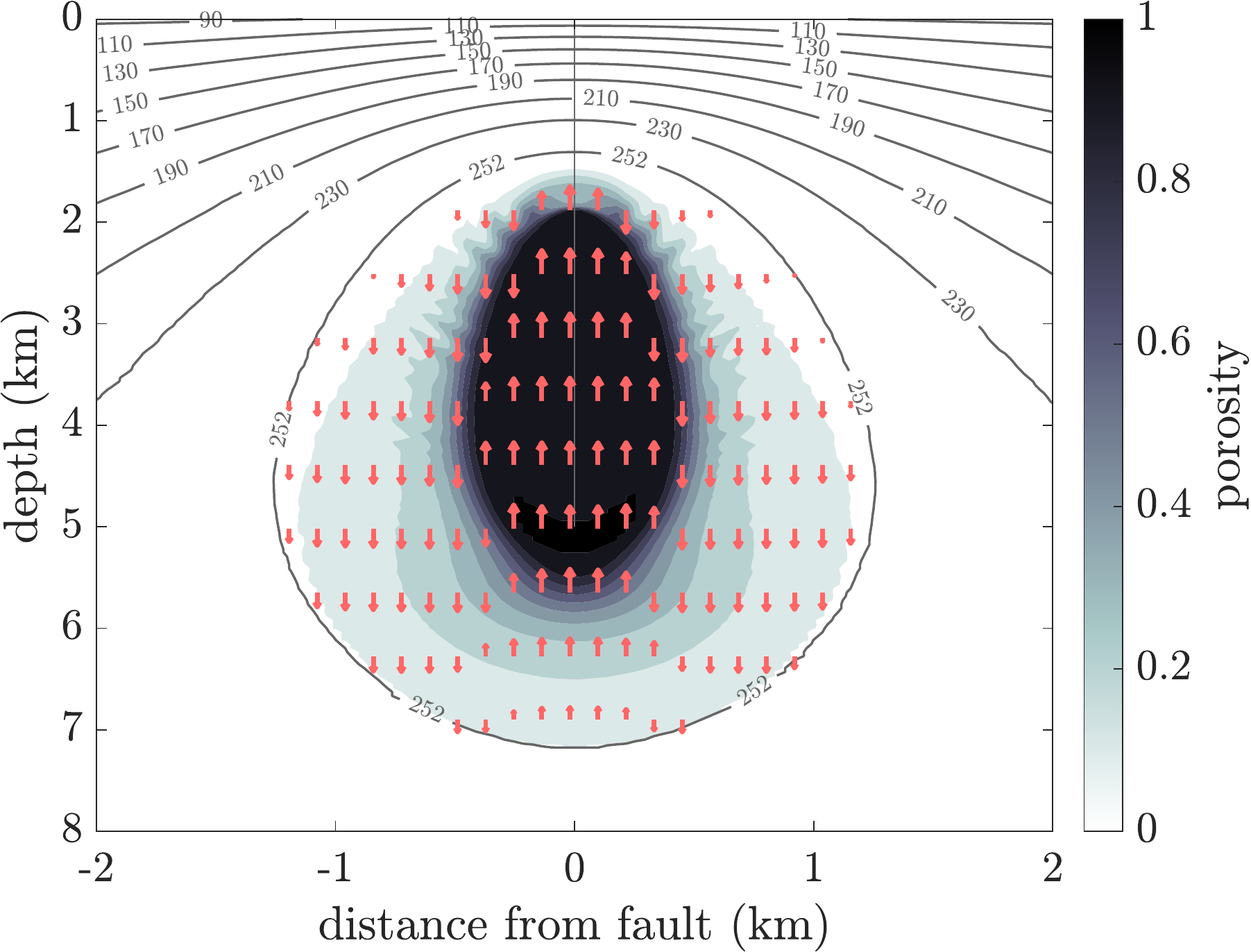}
\caption{Mushy zone fluid reservoir surrounding tiger stripe fractures: (left) Steady interstitial fluid volume and peak melt rate as a function of the shear heat flux into the solid: dimensional on top and nondimensional on bottom. At a critical value of $F\approx 170$, a mushy zone develops around the fracture and this near-surface liquid water can erupt as jets. For larger heat fluxes, the volume of the reservoir and peak melt rate increase. (right) Porosity $\phi$ with superimposed temperature contours from figure \ref{fig:tempcontour} for $F=500$ and the vertical component of the interstitial velocity (logarithmic magnitude, peak on the order of 1 mm/yr, see supplement). The porosity is zero outside the mushy zone, where the temperatures are below the eutectic. The porosity reaches unity along part of the crack. Additional plots for $F=180$, $F=240$, and $F=420$ are shown in the supplement.}
\label{fig:Fvolume}
\end{figure}

The observed flux of ice and vapor from the plume is on the order of $200$~kg/s, which is a volume flux of about $0.2$~m$^3$/s \citep{Han2008}. With the volume for $F=500$, this flux would take about 350 thousand years to deplete the reservoir, if no new melting occurred. The volume calculation for $F=500$ elucidates two important ideas. First, that there is sufficient interstitial brine in the mushy zone surrounding the fracture to account for the plume volume. Second, the volume of material required for the jets is relatively small when compared to the size of the ocean, which is on the order of $20$ million km$^{3}$ \citep{Spe2013}. In other words, it is conceivable that the plumes could be maintained without direct access to a large reservoir, such as a subsurface ocean. Thus, plumes on Enceladus and on other icy satellites may be sourced from smaller liquid reservoirs produced by shear heating along faults driven by tidal motion.

From transient simulations, we determine how the liquid volume approaches steady state, which allows us to calculate the peak liquid volume production rate (blue, right axis, figure \ref{fig:Fvolume}). The peak melt rate estimate for $F=500$ (slip velocity of $u=3.4$ $\mu$m s$^{-1}$, see methods) is consistent with the inferred $\rm H_2O$ flux and sufficient for E-ring particle source estimates \citep{Ing2011}. Although the melt rate is not necessarily equal to the ejection rate, sufficient melt must be available to replenish the liquid source. Here we assume that the dominant impurity is NaCl; the presence of additional impurities with lower eutectic points would likely lead to a larger predicted melt volume and melt production rate, which is something we will explore in future work. Smaller values of the nondimensional heat flux imply lower peak melt rates, but can still maintain large volume fluxes for a long time (e.g. $\sim$30,000~years for $F=200$). In this case, when the reservoir is depleted the friction on the fault would increase and shear heating would build up a new mushy zone, leading to a long-period oscillation in geyser activity, with a period on the order of the depletion time, i.e. $\sim$30,000~years for $F=200$. This could be analogous to the hypothesis that Enceladus' ocean varies in thickness periodically, due to  mismatches in heat output and tidal heat input \citep{Spe2013,Nim2018}. Although these simulations represent an upper bound on the rate of shear heating, this mechanism is sufficient to explain the observed geyser ejection rate and required brine volume.

\section*{Discussion} 
Observations show that the plume brightness varies diurnally with a phase lag \citep{Nim2014,Por2014}, which is consistent with \citet{Hur2007}, i.e. that as normal stresses on the faults change into a more tensile state due to tides, the brightness of the plume also changes. As the cracks open up, more material is able to come up from the fluid reservoir and emanate out as the jets. Here we postulate that the reservoir is not necessarily a subsurface ocean, but could instead be a mushy zone within the ice shell. A connection from the surface to a reservoir is still required, but it is not necessary for the fracture to extend all the way through the ice shell. The lag from the tidal variation observed by \citet{Nim2014} is still consistent with this mechanism so long as the opening of the fractures allows material to emerge, just as in the \citet{Hur2007} model. Moreover, the mushy zone would naturally generate a low-viscosity region around the tiger stripes that \citet{Beh2015} used to explain the phase lag. \citet{Kit2016} proposed that the phase lag pushes the crack faces together causing the fluid to rise turbulently. However, the tidal squeezing implied by their model is difficult to reconcile because the plumes are still active during the full tidal cycle. In any case, if there are salts present in the shell a mushy zone around the crack will likely form in the \citet{Kit2016} case too because of the heat flux on the crack.

The double ridges observed around the tiger stripe fractures on the south pole of Enceladus \citep{Pat2018} and in many places on Europa may provide an additional line of evidence suggesting that shear heating along the fractures produces internal melting. \citet{Cul2022} describe double-ridge formation on icy satellites building on evidence from refreezing in the surface snow of the Greenland Ice Sheet. As liquid water freezes in a near-surface reservoir, it expands and drives the sides of the crack vertically. Similarly, freezing of a liquid-water pocket could propagate a vertical crack, leading to geysers and double ridges \citep{Les2022}. The mushy zone that we propose could be a near-surface water source and the double ridges could be evidence for episodic geysering followed by refreezing dormant periods. Triton is another place where double ridges may be associated with shear heating \citep{Pro2005}.

Within the mushy zone generated by the shear heating mechanism, the source brine is concentrated interstitial liquid produced from the remelting of the ice shell. Thus, it may not be representative of the underlying ocean chemistry (see figure 5 in the supplement). The difference between interstitial brine and the ocean has fundamental implications for how we relate observed plume properties to interior liquid water reservoirs. Cryoconcentrative processes in the shell and ion fractionation during droplet solidification may result in plume compositions that vary significantly from that of the shell or ocean \citep{Buf2021b,Buf2023,Fox2021}. Such uncertainties feed into predictions of ocean geochemistry and habitability based on observed plume compositions. We suggest caution in extrapolating plume properties to interior ocean properties. Without corroborating evidence of a direct ocean-to-surface conduit, the plume may be sourced from an intrashell brine reservoir whose composition and concentration will depend on the geophysical history and dynamics of the hydrologic body and surrounding ice shell.

\section*{Conclusions}
We have described the potential for a mushy source for the geysers of Enceladus, by revisiting the \citet{Nim2007} idea that shear heating along the tiger stripes is responsible for the jet material. Instead of treating the shell as pure ice, we consider a composite of solid ice and salt, which would naturally form when freezing a salty ocean. Using an idealized model, we have shown that shear heating along the tiger stripe fractures can generate a mushy zone at depth, while maintaining cold surface temperatures, in line with \citet{Abr2009}. Within the mushy zone, we find that there is sufficient liquid volume in the interstices to source the volume of observed geyser material and that the melting rate can match the geyser ejection rate. However, the composition of the liquid brine within the mushy zone may be distinct from that of the ocean.

\section*{Methods}
We consider shear heating on a fault with tidally driven back-and-forth strike-slip motion. The shear stress on the fault depends on the friction coefficient $\mu$ and the effective normal stress $N$ (overburden weight $\rho_i g z$ less liquid pore pressure $p_w$, i.e. $N = \rho_i g z-p_w$, where $z$ is the depth below the surface). When melting occurs on the fault, fluid pressure can limit fault stress by lowering the effective normal stress \citep{Ric2006}. Shear heating $\mu N u$ is given by the shear stress $\mu N$ multiplied by the multiplied by the slip velocity $u$ \citep{Nie2008}. Due to model constraints and simplicity, we use a constant overburden stress, rather than a depth-increasing overburden, and do not incorporate the water pressure on the crack face. In other words, we take the rate of shear heating to be the slip velocity $u$ multiplied by the overburden stress, i.e. $\mu \rho_i g h u$, where $\rho_i$ the ice density, $g$ the gravity, and $h$ the crack depth \citep{Nim2002}. In effect, our heat flux into the crack is an upper bound on the total potential shear heating. Sliding beneath the fracture is accommodated by viscous deformation but we do not explicitly account for viscous deformation and thermal weakening \citep[e.g.][]{Mey2018a} in this region. We also consider the heating rate to be temporally constant, i.e. not varying with time due to tides. This is an acceptable simplification because the heat diffusion timescale is much longer than the orbital timescale. Acknowledging these simplifications, we assume that the friction decreases with added melt according to $\mu (1-\phi)$ where the porosity $\phi$ is a variable that is solved for in the simulation. The shear heating along the fault $f_c$ is defined as
\begin{equation}
f_c = \mu (1-\phi) \rho_i g h u .
\label{eqn:fc}
\end{equation}
As the porosity grows, the heat flux will decay due to liquid reducing friction along the fault, reducing the heat production.

Following \citet{Par2020}, we write the nondimensional temperature $\theta$ as 
\begin{equation}\theta = \frac{T-T_e}{\Delta T},~~~ \Delta T = T_{L}(C_i) - T_e\label{eqn:theta}\end{equation}
where $T_e$ is the eutectic temperature and $T_L(C_i)$ is the temperature on the liquidus at the initial salt concentration $C_i$. Here we use the idealized liquidus equation
\begin{equation}
T_L = T_m - m C,
\label{eqn:liquidus}
\end{equation}
where $T_m=273.15$ $^\circ$C is the pure ice melting temperature and $C$ is the salt concentration (see figure~\ref{fig:shellschematic}). We scale heights with a characteristic shell thickness $H = 10$ km and use NaCl parameters with ${m=0.0913}$~K/ppt and an initial concentration of $C_i = 35$ ppt, similar to Earth's ocean and in the range of observations for Enceladus' plumes \citep{Pos2009}. A full table of parameters is given in the supplemental information. Our choice of parameters, such as the crack depth and composition, affects the results quantitatively but not qualitatively.

Considering the scales for depth and temperature, we write the constant part of the nondimensional heat flux into the fracture $F$ as
\begin{equation}
F = \frac{\mu \rho_i g h u H}{k_o \Delta T},
\end{equation}
where $k_o$ is the thermal conductivity of saltwater. The full nondimensional heat flux (i.e. equation \ref{eqn:fc} divided by $k_o \Delta T/H$) is therefore given as 
\begin{equation}
F_c = F (1-\phi).
\end{equation}
In our simulations, $F$ is the primary control parameter. Taking the parameter values of slip velocity ${u=3.4\times10^{-6}}$ m s$^{-1}$, coefficient of friction $\mu=0.3$ \citep{Gol2013}, ice density $\rho_i=940$ kg m$^{-3}$ \citep{Gol1998}, gravity $g=0.113$ m s$^{-2}$, and crack depth $h=5$ km, we find that $\mu \rho_i g h u=0.54$ W m$^{-2}$ and $F = 500$ (see table in the supplement). Here the value for $u$ is in the range described by \citet{Nim2007} and produces the value for $F$ that we use in figures \ref{fig:tempcontour} and \ref{fig:Fvolume}.

SOFTBALL (SOlidification, Flow, and Thermodynamics in a Binary ALLoy) is an open-source reactive transport code that we use to solve for temperature, porosity, and salt content as well as fluid flow through the interstices of evolving mushy zones \citep{Par2020,Ada2021}. The set up is shown schematically in figure \ref{fig:tempcontour}. We apply a linearized radiative condition at the top of the ice shell, and insulating conditions on the sides of the domain. We also set the temperature at the eutectic at the bottom of the domain to simulate the top of the ice-ocean interfacial mushy zone. We run the simulations in two dimensions, treating the out-of-plane dimension as invariant along the tiger stripe fractures. The heat flux $f_c$ is implemented via a Gaussian heat source localized at the center of the domain (width $\lesssim$500 m) and a hyperbolic function in the vertical direction (decay away from the crack tip $\lesssim$1000 m). We initialize the SOFTBALL simulations with a steady state conductive temperature profile and a uniformly distributed initial salt concentration ($C_i=35$ ppt). We then run the simulation to steady state ($\sim$4 million years; figure 2 in the supplement). Additional details on the simulations are included in the supplement. 

\section*{Acknowledgements}
We thank Mark Fox-Powell, Alan Rempel, Aleah Sommers, and John Spencer for insightful conversations. This work was partially supported by NASA 21-EPSCoR2021-0024, NASA 20-SSW20-0049, ARO 78811EG, and the William H. Neukom Institute for Computational Science at Dartmouth College. The authors have no conflict of interest. We did not use any raw observational data in this paper beyond the values from cited papers and the SOFTBALL input files can be found here $<$link inserted in proofs$>$. 

\bibliographystyle{plainnat}
\bibliography{Glib}

\begin{thebibliography}{55}
\providecommand{\natexlab}[1]{#1}
\providecommand{\url}[1]{\texttt{#1}}
\expandafter\ifx\csname urlstyle\endcsname\relax
  \providecommand{\doi}[1]{doi: #1}\else
  \providecommand{\doi}{doi: \begingroup \urlstyle{rm}\Url}\fi

\bibitem[Abramov and Spencer(2009)]{Abr2009}
O.~Abramov and J.~R. Spencer.
\newblock Endogenic heat from {E}nceladus' south polar fractures: New
  observations, and models of conductive surface heating.
\newblock \emph{Icarus}, 199\penalty0 (1):\penalty0 189 -- 196, 2009.
\newblock \doi{10.1016/j.icarus.2008.07.016}.

\bibitem[Adams et~al.(2021)Adams, Colella, Graves, Johnson, Keen, Ligocki,
  Martin, McCorquodale, Modiano, Schwartz, Sternberg, and Straalen]{Ada2021}
M.~Adams, P.~Colella, D.~T. Graves, J.~N. Johnson, N.~D. Keen, T.~J. Ligocki,
  D.~F. Martin, P.~W. McCorquodale, D.~Modiano, P.~O. Schwartz, T.~D.
  Sternberg, and B.~Van Straalen.
\newblock Chombo software package for {AMR} applications - design document.
\newblock Technical report, Lawrence Berkeley National Laboratory, LBNL-6616E,
  2021.

\bibitem[B{\v{e}}hounkov{\'a} et~al.(2015)B{\v{e}}hounkov{\'a}, Tobie,
  {\v{C}}adek, Choblet, Porco, and Nimmo]{Beh2015}
M.~B{\v{e}}hounkov{\'a}, G.~Tobie, O.~{\v{C}}adek, G.~Choblet, C.~Porco, and
  F.~Nimmo.
\newblock {Timing of water plume eruptions on Enceladus explained by interior
  viscosity structure}.
\newblock \emph{Nat. Geosci.}, 8\penalty0 (8):\penalty0 601--604, 2015.
\newblock \doi{10.1038/ngeo2475}.

\bibitem[Bender et~al.(1997)Bender, Sowers, and Brook]{Ben1997}
M.~Bender, T.~Sowers, and Ed. Brook.
\newblock Gases in ice cores.
\newblock \emph{Proc. Natl. Acad. Sci. U.S.A.}, 94\penalty0 (16):\penalty0
  8343--8349, 1997.
\newblock \doi{10.1073/pnas.94.16.8343}.

\bibitem[Bombosch and Jenkins(1995)]{Bom1995}
A.~Bombosch and A.~Jenkins.
\newblock Modeling the formation and deposition of frazil ice beneath
  {Filchner-Ronne Ice Shelf}.
\newblock \emph{J. Geophys. Res.}, 100\penalty0 (C4):\penalty0 6983--6992,
  1995.
\newblock \doi{10.1029/94JC03224}.

\bibitem[Boury et~al.(2021)Boury, Meyer, Vasil, and Wells]{Bou2021}
S.~Boury, C.~R. Meyer, G.~M. Vasil, and A.~J. Wells.
\newblock Convection in a mushy layer along a vertical heated wall.
\newblock \emph{J. Fluid Mech.}, 926:\penalty0 A33, 2021.
\newblock \doi{10.1017/jfm.2021.742}.

\bibitem[Buffo et~al.(2018)Buffo, Schmidt, and Huber]{Buf2018}
J.~J. Buffo, B.~E. Schmidt, and C.~Huber.
\newblock Multiphase reactive transport and platelet ice accretion in the sea
  ice of {McMurdo Sound, Antarctica}.
\newblock \emph{J. Geophys. Res.}, 123\penalty0 (1):\penalty0 324--345, 2018.
\newblock \doi{10.1002/2017JC013345}.

\bibitem[Buffo et~al.(2021{\natexlab{a}})Buffo, Meyer, and Parkinson]{Buf2021b}
J.~J. Buffo, C.~R. Meyer, and J.~R.~G. Parkinson.
\newblock Dynamics of a solidifying icy satellite shell.
\newblock \emph{J. Geophys. Res.}, 126\penalty0 (5):\penalty0 e2020JE006741,
  2021{\natexlab{a}}.
\newblock \doi{10.1029/2020JE006741}.

\bibitem[Buffo et~al.(2021{\natexlab{b}})Buffo, Schmidt, Huber, and
  Meyer]{Buf2021a}
J.~J. Buffo, B.~E. Schmidt, C.~Huber, and C.~R. Meyer.
\newblock Characterizing the ice-ocean interface of icy worlds: A theoretical
  approach.
\newblock \emph{Icarus}, 360:\penalty0 114318, 2021{\natexlab{b}}.
\newblock ISSN 0019--1035.
\newblock \doi{10.1016/j.icarus.2021.114318}.

\bibitem[Buffo et~al.(2023)Buffo, Meyer, Chivers, Walker, Huber, and
  Schmidt]{Buf2023}
J.~J. Buffo, C.~R. Meyer, C.~J. Chivers, C.~C. Walker, C.~Huber, and B.~E.
  Schmidt.
\newblock Geometry of freezing impacts ice composition: Implications for icy
  satellites.
\newblock \emph{J. Geophys. Res.}, 128\penalty0 (3):\penalty0 e2022JE007389,
  2023.
\newblock \doi{10.1029/2022JE007389}.

\bibitem[Culberg et~al.(2022)Culberg, Schroeder, Walker, and
  Steinbr\:{u}gge]{Cul2022}
R.~Culberg, D.~M. Schroeder, M.~Walker, and G.~Steinbr\:{u}gge.
\newblock {Double ridge formation over shallow water sills on Jupiter’s moon
  Europa}.
\newblock \emph{Nat. Comm.}, 13\penalty0 (2007):\penalty0 2041--1723, 2022.
\newblock \doi{10.1038/s41467-022-29458-3}.

\bibitem[Fowler(1985)]{Fow1985}
A.~C. Fowler.
\newblock The formation of freckles in binary alloys.
\newblock \emph{IMA J. Appl. Math.}, 35\penalty0 (2):\penalty0 159--174, 1985.
\newblock \doi{10.1093/imamat/35.2.159}.

\bibitem[Fox-Powell and Cousins(2021)]{Fox2021}
M.~G. Fox-Powell and C.~R. Cousins.
\newblock Partitioning of crystalline and amorphous phases during freezing of
  simulated enceladus ocean fluids.
\newblock \emph{J. Geophys. Res.}, 126\penalty0 (1):\penalty0 e2020JE006628,
  2021.
\newblock \doi{10.1029/2020JE006628}.

\bibitem[Golden et~al.(1998)Golden, Ackley, and Lytle]{Gol1998}
K.~M. Golden, S.~F. Ackley, and V.~I. Lytle.
\newblock The percolation phase transition in sea ice.
\newblock \emph{Science}, 282\penalty0 (5397):\penalty0 2238--2241, 1998.
\newblock \doi{10.1126/science.282.5397.2238}.

\bibitem[Golding et~al.(2010)Golding, Schulson, and Renshaw]{Gol2010}
N.~Golding, E.~M. Schulson, and C.~E. Renshaw.
\newblock Shear faulting and localized heating in ice: The influence of
  confinement.
\newblock \emph{Acta Mater.}, 58\penalty0 (15):\penalty0 5043 -- 5056, 2010.
\newblock \doi{10.1016/j.actamat.2010.05.040}.

\bibitem[Golding et~al.(2013)Golding, Burks, Lucas, Fortt, Snyder, and
  Schulson]{Gol2013}
N.~Golding, C.~E. Burks, K.~N. Lucas, A.~L. Fortt, S.~A. Snyder, and E.~M.
  Schulson.
\newblock Mechanical properties of the ice i--magnesium sulfate eutectic: A
  comparison with freshwater ice in reference to {E}uropa.
\newblock \emph{Icarus}, 225\penalty0 (1):\penalty0 248--256, 2013.
\newblock \doi{10.1016/j.icarus.2013.02.037}.

\bibitem[Hammond(2020)]{Hamm2020}
N.~P. Hammond.
\newblock Estimating the magnitude of cyclic slip on strike-slip faults on
  {E}uropa.
\newblock \emph{J. Geophys. Res.}, 125\penalty0 (7):\penalty0 e2019JE006170,
  2020.
\newblock \doi{10.1029/2019JE006170}.

\bibitem[{Hammond} et~al.(2018){Hammond}, {Parmentier}, and {Barr}]{Ham2018}
N.~P. {Hammond}, E.~M. {Parmentier}, and A.~C. {Barr}.
\newblock Compaction and melt transport in ammonia-rich ice shells:
  Implications for the evolution of {T}riton.
\newblock \emph{J. Geophys. Res.}, 123\penalty0 (12):\penalty0 3105--3118,
  2018.
\newblock \doi{10.1029/2018JE005781}.

\bibitem[Hansen et~al.(2008)Hansen, Esposito, Stewart, Meinke, Wallis, Colwell,
  Hendrix, Larsen, Pryor, and Tian]{Han2008}
C.~J. Hansen, L.~W. Esposito, A.~I.~F. Stewart, B.~Meinke, B.~Wallis, J.~E.
  Colwell, A.~R. Hendrix, K.~Larsen, W.~Pryor, and F.~Tian.
\newblock Water vapour jets inside the plume of gas leaving enceladus.
\newblock \emph{Nature}, 456\penalty0 (7221):\penalty0 477--479, 2008.
\newblock \doi{10.1038/nature07542}.

\bibitem[Hedman et~al.(2013)Hedman, Gosmeyer, Nicholson, Sotin, Brown, Clark,
  Baines, Buratti, and Showalter]{Hed2013}
M.~M. Hedman, C.~M. Gosmeyer, P.~D. Nicholson, C.~Sotin, R.~H. Brown, R.~N.
  Clark, K.~H. Baines, B.~J. Buratti, and M.~R. Showalter.
\newblock An observed correlation between plume activity and tidal stresses on
  {E}nceladus.
\newblock \emph{Nature}, 500:\penalty0 182, 2013.
\newblock \doi{10.1038/nature12371}.

\bibitem[{Hsu} et~al.(2015){Hsu}, {Postberg}, {Sekine}, {Shibuya}, {Kempf},
  {Hor{\'a}nyi}, {Juh{\'a}sz}, {Altobelli}, {Suzuki}, {Masaki}, {Kuwatani},
  {Tachibana}, {Sirono}, {Moragas-Klostermeyer}, and {Srama}]{Hsu2015}
H.-W. {Hsu}, Frank {Postberg}, Y.~{Sekine}, T.~{Shibuya}, S.~{Kempf},
  M.~{Hor{\'a}nyi}, A.~{Juh{\'a}sz}, N.~{Altobelli}, K.~{Suzuki}, Y.~{Masaki},
  T.~{Kuwatani}, S.~{Tachibana}, S.-I. {Sirono}, G.~{Moragas-Klostermeyer}, and
  R.~{Srama}.
\newblock {Ongoing hydrothermal activities within Enceladus}.
\newblock \emph{Nature}, 519\penalty0 (7542):\penalty0 207--210, 2015.
\newblock \doi{10.1038/nature14262}.

\bibitem[Hunke et~al.(2011)Hunke, Notz, Turner, and Vancoppenolle]{Hun2011}
E.~C. Hunke, D.~Notz, A.~K. Turner, and M.~Vancoppenolle.
\newblock The multiphase physics of sea ice: a review for model developers.
\newblock \emph{Cryosphere}, 5\penalty0 (4):\penalty0 989--1009, 2011.
\newblock \doi{10.5194/tc-5-989-2011}.

\bibitem[{Hurford} et~al.(2007){Hurford}, {Helfenstein}, {Hoppa}, {Greenberg},
  and {Bills}]{Hur2007}
T.~A. {Hurford}, P.~{Helfenstein}, G.~V. {Hoppa}, R.~{Greenberg}, and B.~G.
  {Bills}.
\newblock {Eruptions arising from tidally controlled periodic openings of rifts
  on {E}nceladus}.
\newblock \emph{Nature}, 447:\penalty0 292--294, 2007.
\newblock \doi{10.1038/nature05821}.

\bibitem[Ingersoll and Ewald(2011)]{Ing2011}
A.~P. Ingersoll and S.~P. Ewald.
\newblock {Total particulate mass in Enceladus plumes and mass of {Saturn}’s
  E ring inferred from Cassini ISS images}.
\newblock \emph{Icarus}, 216\penalty0 (2):\penalty0 492--506, 2011.
\newblock ISSN 0019-1035.
\newblock \doi{10.1016/j.icarus.2011.09.018}.

\bibitem[{Ingersoll} and {Ewald}(2017)]{Ing2017}
A.~P. {Ingersoll} and S.~P. {Ewald}.
\newblock {Decadal timescale variability of the {E}nceladus plumes inferred
  from {C}assini images}.
\newblock \emph{Icarus}, 282:\penalty0 260--275, 2017.
\newblock \doi{10.1016/j.icarus.2016.09.018}.

\bibitem[Ingersoll and Nakajima(2016)]{Ing2016}
A.~P. Ingersoll and M.~Nakajima.
\newblock Controlled boiling on {E}nceladus. 2. model of the liquid-filled
  cracks.
\newblock \emph{Icarus}, 272:\penalty0 319 -- 326, 2016.
\newblock \doi{10.1016/j.icarus.2015.12.040}.

\bibitem[{Kalousov{\'a}} et~al.(2016){Kalousov{\'a}}, {Sou{\v c}ek}, {Tobie},
  {Choblet}, and {{\v C}adek}]{Kal2016}
K.~{Kalousov{\'a}}, O.~{Sou{\v c}ek}, G.~{Tobie}, G.~{Choblet}, and O.~{{\v
  C}adek}.
\newblock {Water generation and transport below {E}uropa's strike-slip faults}.
\newblock \emph{J. Geophys. Res.}, 121:\penalty0 2444--2462, 2016.
\newblock \doi{10.1002/2016JE005188}.

\bibitem[Katz(2008)]{Kat2008a}
R.~F. Katz.
\newblock Magma dynamics with the enthalpy method: Benchmark solutions and
  magmatic focusing at mid-ocean ridges.
\newblock \emph{J. Petrol.}, 49\penalty0 (12):\penalty0 2099--2121, 12 2008.
\newblock ISSN 0022-3530.
\newblock \doi{10.1093/petrology/egn058}.

\bibitem[Kite and Rubin(2016)]{Kit2016}
E.~S. Kite and A.~M. Rubin.
\newblock Sustained eruptions on {E}nceladus explained by turbulent dissipation
  in tiger stripes.
\newblock \emph{Proc. Natl. Acad. Sci.}, 113\penalty0 (15):\penalty0
  3972--3975, 2016.
\newblock \doi{10.1073/pnas.1520507113}.

\bibitem[Lesage et~al.(2022)Lesage, Massol, Howell, and Schmidt]{Les2022}
E.~Lesage, H.~Massol, S.~M. Howell, and F.~Schmidt.
\newblock Simulation of freezing cryomagma reservoirs in viscoelastic ice
  shells.
\newblock \emph{Planet. Sci. J.}, 3\penalty0 (7):\penalty0 170, 2022.
\newblock \doi{10.3847/PSJ/ac75bf}.

\bibitem[Lister(1990)]{Lis1990}
J.~R. Lister.
\newblock Buoyancy-driven fluid fracture: similarity solutions for the
  horizontal and vertical propagation of fluid-filled cracks.
\newblock \emph{J. Fluid Mech.}, 217:\penalty0 213--239, 1990.
\newblock \doi{10.1017/S0022112090000696}.

\bibitem[Meyer and Minchew(2018)]{Mey2018a}
C.~R. Meyer and B.~M. Minchew.
\newblock Temperate ice in the shear margins of the {Antarctic} ice sheet:
  Controlling processes and preliminary locations.
\newblock \emph{Earth Planet. Sci. Lett.}, 498:\penalty0 17 -- 26, 2018.
\newblock ISSN 0012-821X.
\newblock \doi{10.1016/j.epsl.2018.06.028}.

\bibitem[Nakajima and Ingersoll(2016)]{Nak2016}
M.~Nakajima and A.~P. Ingersoll.
\newblock Controlled boiling on {E}nceladus. 1. model of the vapor-driven jets.
\newblock \emph{Icarus}, 272:\penalty0 309 -- 318, 2016.
\newblock \doi{https://doi.org/10.1016/j.icarus.2016.02.027}.

\bibitem[Nielsen et~al.(2008)Nielsen, Di~Toro, Hirose, and Shimamoto]{Nie2008}
S.~Nielsen, G.~Di~Toro, T.~Hirose, and T.~Shimamoto.
\newblock Frictional melt and seismic slip.
\newblock \emph{J. Geophys. Res.}, 113\penalty0 (B1), 2008.
\newblock \doi{10.1029/2007JB005122}.

\bibitem[Nimmo and Gaidos(2002)]{Nim2002}
F.~Nimmo and E.~Gaidos.
\newblock Strike-slip motion and double ridge formation on {E}uropa.
\newblock \emph{J. Geophys. Res.}, 107\penalty0 (E4):\penalty0 5--1--5--8,
  2002.
\newblock \doi{10.1029/2000JE001476}.

\bibitem[Nimmo et~al.(2007)Nimmo, Spencer, Pappalardo, and Mullen]{Nim2007}
F.~Nimmo, J.~R. Spencer, R.~T. Pappalardo, and M.~E. Mullen.
\newblock Shear heating as the origin of the plumes and heat flux on
  {E}nceladus.
\newblock \emph{Nature}, 447\penalty0 (7142):\penalty0 289, 2007.
\newblock \doi{10.1038/nature05783}.

\bibitem[Nimmo et~al.(2014)Nimmo, Porco, and Mitchell]{Nim2014}
F.~Nimmo, C.~Porco, and C.~Mitchell.
\newblock Tidally modulated eruptions on {E}nceladus: Cassini {ISS}
  observations and models.
\newblock \emph{Astron. J.}, 148\penalty0 (3):\penalty0 46, 2014.
\newblock \doi{10.1088/0004-6256/148/3/46}.

\bibitem[Nimmo et~al.(2018)Nimmo, Barr, B\v{e}hounkov\'{a}, and
  McKinnon]{Nim2018}
F.~Nimmo, A.~C. Barr, M.~B\v{e}hounkov\'{a}, and W.~B. McKinnon.
\newblock The thermal and orbital evolution of {E}nceladus: Observational
  constraints and models.
\newblock In P.~M. Schenk, R.~N. Clark, C.~J.~A. Howett, A.~J. Verbiscer, and
  J.~H. Waite, editors, \emph{Enceladus and the {I}cy {M}oons of {S}aturn},
  pages 79--94. University of Arizona Press, 2018.
\newblock ISBN 9780816537075.
\newblock \doi{10.2307/j.ctv65sw2b.13}.

\bibitem[Parkinson et~al.(2020)Parkinson, Martin, Wells, and Katz]{Par2020}
J.~R.~G. Parkinson, D.~F. Martin, A.~J. Wells, and R.~F. Katz.
\newblock Modelling binary alloy solidification with adaptive mesh refinement.
\newblock \emph{J. Comput. Phys.}, 5:\penalty0 100043, 2020.
\newblock \doi{10.1016/j.jcpx.2019.100043}.

\bibitem[{Patterson} et~al.(2018){Patterson}, {Kattenhorn}, {Helfenstein},
  {Collins}, and {Pappalardo}]{Pat2018}
G.~W. {Patterson}, S.~A. {Kattenhorn}, P.~{Helfenstein}, G.~C. {Collins}, and
  R.~T. {Pappalardo}.
\newblock The interior of {E}nceladus.
\newblock In P.~M. Schenk, R.~N. Clark, C.~J.~A. Howett, A.~J. Verbiscer, and
  J.~H. Waite, editors, \emph{Enceladus and the {I}cy {M}oons of {S}aturn},
  page~95. University of Arizona Press, 2018.
\newblock ISBN 9780816537075.
\newblock \doi{10.2307/j.ctv65sw2b.12}.

\bibitem[Patterson and Saltzman(2021)]{Pat2021}
J.~D. Patterson and E.~S. Saltzman.
\newblock Diffusivity and solubility of {H2} in ice {Ih}: Implications for the
  behavior of {H2} in polar ice.
\newblock \emph{J. Geophys. Res.}, 126\penalty0 (10):\penalty0 e2020JD033840,
  2021.
\newblock \doi{https://doi.org/10.1029/2020JD033840}.

\bibitem[Porco et~al.(2006)Porco, Helfenstein, Thomas, Ingersoll, Wisdom, West,
  Neukum, Denk, Wagner, Roatsch, et~al.]{Por2006}
C.~Porco, P.~Helfenstein, P.~C. Thomas, A.~P. Ingersoll, J.~Wisdom, R.~West,
  G.~Neukum, T.~Denk, R.~Wagner, T.~Roatsch, et~al.
\newblock Cassini observes the active south pole of {E}nceladus.
\newblock \emph{Science}, 311\penalty0 (5766):\penalty0 1393--1401, 2006.
\newblock \doi{10.1126/science.1123013}.

\bibitem[Porco et~al.(2014)Porco, DiNino, and Nimmo]{Por2014}
C.~Porco, D.~DiNino, and F.~Nimmo.
\newblock How the geysers, tidal stresses, and thermal emission across the
  south polar terrain of {Enceladus} are related.
\newblock \emph{Astron. J.}, 148\penalty0 (3):\penalty0 45, 2014.
\newblock \doi{10.1088/0004-6256/148/3/45}.

\bibitem[Postberg et~al.(2009)Postberg, {Kempf}, {Schmidt}, {Brilliantov},
  {Beinsen}, {Abel}, {Buck}, and {Srama}]{Pos2009}
F.~Postberg, S.~{Kempf}, J.~{Schmidt}, N.~{Brilliantov}, A.~{Beinsen},
  B.~{Abel}, U.~{Buck}, and R.~{Srama}.
\newblock {Sodium salts in {E}-ring ice grains from an ocean below the surface
  of {E}nceladus}.
\newblock \emph{Nature}, 459:\penalty0 1098--1101, 2009.
\newblock \doi{10.1038/nature08046}.

\bibitem[Postberg et~al.(2011)Postberg, Schmidt, Hillier, Kempf, and
  Srama]{Pos2011}
F.~Postberg, J.~Schmidt, J.~Hillier, S.~Kempf, and R.~Srama.
\newblock A salt-water reservoir as the source of a compositionally stratified
  plume on {E}nceladus.
\newblock \emph{Nature}, 474\penalty0 (7353):\penalty0 620, 2011.
\newblock \doi{10.1038/nature10175}.

\bibitem[Postberg et~al.(2018)Postberg, Khawaja, Abel, Choblet, Glein,
  Gudipati, Henderson, Hsu, Kempf, Klenner, Moragas-Klostermeyer, Magee,
  N{\"o}lle, Perry, Reviol, Schmidt, Srama, Stolz, Tobie, Trieloff, and
  Waite]{Pos2018}
F.~Postberg, N.~Khawaja, B.~Abel, G.~Choblet, C.~R. Glein, M.~S. Gudipati,
  B.~L. Henderson, H.-W. Hsu, S.~Kempf, F.~Klenner, G.~Moragas-Klostermeyer,
  B.~Magee, L.~N{\"o}lle, M.~Perry, R.~Reviol, J.~Schmidt, R.~Srama, F.~Stolz,
  G.~Tobie, M.~Trieloff, and J.~H. Waite.
\newblock Macromolecular organic compounds from the depths of {Enceladus}.
\newblock \emph{Nature}, 558\penalty0 (7711):\penalty0 564--568, 2018.
\newblock \doi{10.1038/s41586-018-0246-4}.

\bibitem[Prockter et~al.(2005)Prockter, Nimmo, and Pappalardo]{Pro2005}
L.~M. Prockter, F.~Nimmo, and R.~T. Pappalardo.
\newblock A shear heating origin for ridges on {Triton}.
\newblock \emph{Geophys. Res. Lett.}, 32\penalty0 (14), 2005.
\newblock \doi{https://doi.org/10.1029/2005GL022832}.

\bibitem[Rice(2006)]{Ric2006}
J.~R. Rice.
\newblock Heating and weakening of faults during earthquake slip.
\newblock \emph{J. Geophys. Res.}, 111\penalty0 (B5), 2006.
\newblock \doi{10.1029/2005JB004006}.

\bibitem[Spencer and Nimmo(2013)]{Spe2013}
J.~R. Spencer and F.~Nimmo.
\newblock Enceladus: An active ice world in the {Saturn} system.
\newblock \emph{Annu. Rev. Earth Planet. Sci.}, 41\penalty0 (1):\penalty0
  693--717, 2013.
\newblock \doi{10.1146/annurev-earth-050212-124025}.

\bibitem[Stevenson(1996)]{Ste1996}
D.~J. Stevenson.
\newblock Heterogeneous tidal deformation and geysers on {Europa}.
\newblock In \emph{Europa Ocean Conference}, number~5 in Capistrano conference,
  pages 69--70, San Juan Capistrano, Calif., 1996. San Juan Capistrano Res.
  Inst.

\bibitem[Thomas et~al.(2016)Thomas, Tajeddine, Tiscareno, Burns, Joseph,
  Loredo, Helfenstein, and Porco]{Tho2016}
P.~C. Thomas, R.~Tajeddine, M.~S. Tiscareno, J.~A. Burns, J.~Joseph, T.~J.
  Loredo, P.~Helfenstein, and C.~Porco.
\newblock Enceladus's measured physical libration requires a global subsurface
  ocean.
\newblock \emph{Icarus}, 264:\penalty0 37 -- 47, 2016.
\newblock \doi{10.1016/j.icarus.2015.08.037}.

\bibitem[Trumbo et~al.(2019)Trumbo, Brown, and Hand]{Tru2019}
S.~K. Trumbo, M.~E. Brown, and K.~P. Hand.
\newblock Sodium chloride on the surface of {Europa}.
\newblock \emph{Sci. Adv.}, 5\penalty0 (6):\penalty0 eaaw7123, 2019.
\newblock \doi{10.1126/sciadv.aaw7123}.

\bibitem[Waite et~al.(2006)Waite, Combi, Ip, Cravens, McNutt, Kasprzak, Yelle,
  Luhmann, Niemann, Gell, Magee, Fletcher, Lunine, and Tseng]{Wai2006}
J.~H. Waite, M.~R. Combi, W.-H. Ip, T.~E. Cravens, R.~L. McNutt, W.~Kasprzak,
  R.~Yelle, J.~Luhmann, H.~Niemann, D.~Gell, B.~Magee, G.~Fletcher, J.~Lunine,
  and W.-L. Tseng.
\newblock {Cassini Ion and Neutral Mass Spectrometer: E}nceladus plume
  composition and structure.
\newblock \emph{Science}, 311\penalty0 (5766):\penalty0 1419--1422, 2006.
\newblock \doi{10.1126/science.1121290}.

\bibitem[Waite et~al.(2017)Waite, Glein, Perryman, Teolis, Magee, Miller,
  Grimes, Perry, Miller, Bouquet, Lunine, Brockwell, and Bolton]{Wai2017}
J.~H. Waite, C.~R. Glein, R.~S. Perryman, B.~D. Teolis, B.~A. Magee, G.~Miller,
  J.~Grimes, M.~E. Perry, K.~E. Miller, A.~Bouquet, J.~I. Lunine, T.~Brockwell,
  and S.~J. Bolton.
\newblock Cassini finds molecular hydrogen in the {Enceladus} plume: evidence
  for hydrothermal processes.
\newblock \emph{Science}, 356\penalty0 (6334):\penalty0 155--159, 2017.
\newblock \doi{10.1126/science.aai8703}.

\bibitem[Worster(2000)]{Wor2000}
M.~G. Worster.
\newblock Solidification of fluids.
\newblock In G.~K. Batchelor, H.~K. Moffatt, and M.~G. Worster, editors,
  \emph{Perspectives in Fluid Dynamics}, chapter~8, pages 393--444. Cambridge
  University Press, 2000.

\end{thebibliography}


\begin{thebibliography}{37}
\providecommand{\natexlab}[1]{#1}
\providecommand{\url}[1]{\texttt{#1}}
\expandafter\ifx\csname urlstyle\endcsname\relax
  \providecommand{\doi}[1]{doi: #1}\else
  \providecommand{\doi}{doi: \begingroup \urlstyle{rm}\Url}\fi

\bibitem[Abramov and Spencer(2009)]{Abr2009}
O.~Abramov and J.~R. Spencer.
\newblock Endogenic heat from {E}nceladus' south polar fractures: New
  observations, and models of conductive surface heating.
\newblock \emph{Icarus}, 199\penalty0 (1):\penalty0 189 -- 196, 2009.
\newblock \doi{10.1016/j.icarus.2008.07.016}.

\bibitem[Boury et~al.(2021)Boury, Meyer, Vasil, and Wells]{Bou2021}
S.~Boury, C.~R. Meyer, G.~M. Vasil, and A.~J. Wells.
\newblock Convection in a mushy layer along a vertical heated wall.
\newblock \emph{J. Fluid Mech.}, 926:\penalty0 A33, 2021.
\newblock \doi{10.1017/jfm.2021.742}.

\bibitem[Broberg(1999)]{Bro1999}
K.~B. Broberg.
\newblock \emph{Cracks and fracture}.
\newblock Academic Press, 1999.

\bibitem[Buffo et~al.(2018)Buffo, Schmidt, and Huber]{Buf2018}
J.~J. Buffo, B.~E. Schmidt, and C.~Huber.
\newblock Multiphase reactive transport and platelet ice accretion in the sea
  ice of {McMurdo Sound, Antarctica}.
\newblock \emph{J. Geophys. Res.}, 123\penalty0 (1):\penalty0 324--345, 2018.
\newblock \doi{10.1002/2017JC013345}.

\bibitem[Buffo et~al.(2021)Buffo, Meyer, and Parkinson]{Buf2021b}
J.~J. Buffo, C.~R. Meyer, and J.~R.~G. Parkinson.
\newblock Dynamics of a solidifying icy satellite shell.
\newblock \emph{J. Geophys. Res.}, 126\penalty0 (5):\penalty0 e2020JE006741,
  2021.
\newblock \doi{10.1029/2020JE006741}.

\bibitem[Buffo et~al.(2023)Buffo, Meyer, Chivers, Walker, Huber, and
  Schmidt]{Buf2023}
J.~J. Buffo, C.~R. Meyer, C.~J. Chivers, C.~C. Walker, C.~Huber, and B.~E.
  Schmidt.
\newblock Geometry of freezing impacts ice composition: Implications for icy
  satellites.
\newblock \emph{J. Geophys. Res.}, 128\penalty0 (3):\penalty0 e2022JE007389,
  2023.
\newblock \doi{10.1029/2022JE007389}.

\bibitem[{Crawford} and {Stevenson}(1988)]{Cra1988}
G.~D. {Crawford} and D.~J. {Stevenson}.
\newblock {Gas-driven water volcanism in the resurfacing of {E}uropa}.
\newblock \emph{Icarus}, 73:\penalty0 66--79, 1988.
\newblock \doi{10.1016/0019-1035(88)90085-1}.

\bibitem[Hammond(2020)]{Hamm2020}
N.~P. Hammond.
\newblock Estimating the magnitude of cyclic slip on strike-slip faults on
  {E}uropa.
\newblock \emph{J. Geophys. Res.}, 125\penalty0 (7):\penalty0 e2019JE006170,
  2020.
\newblock \doi{10.1029/2019JE006170}.

\bibitem[Hansen et~al.(2006)Hansen, Esposito, Stewart, Colwell, Hendrix, Pryor,
  Shemansky, and West]{Han2006}
C.~J. Hansen, L.~Esposito, A.~I.~F Stewart, J.~Colwell, A.~Hendrix, W.~Pryor,
  D.~Shemansky, and R.~West.
\newblock Enceladus' water vapor plume.
\newblock \emph{Science}, 311\penalty0 (5766):\penalty0 1422--1425, 2006.
\newblock \doi{10.1126/science.1121254}.

\bibitem[Hansen et~al.(2008)Hansen, Esposito, Stewart, Meinke, Wallis, Colwell,
  Hendrix, Larsen, Pryor, and Tian]{Han2008}
C.~J. Hansen, L.~W. Esposito, A.~I.~F. Stewart, B.~Meinke, B.~Wallis, J.~E.
  Colwell, A.~R. Hendrix, K.~Larsen, W.~Pryor, and F.~Tian.
\newblock Water vapour jets inside the plume of gas leaving enceladus.
\newblock \emph{Nature}, 456\penalty0 (7221):\penalty0 477--479, 2008.
\newblock \doi{10.1038/nature07542}.

\bibitem[Hemingway et~al.(2018)Hemingway, Iess, Tajeddine, and Tobie]{Hem2018}
D.~Hemingway, L.~Iess, R.~Tajeddine, and G.~Tobie.
\newblock The interior of {E}nceladus.
\newblock In P.~M. Schenk, R.~N. Clark, C.~J.~A. Howett, A.~J. Verbiscer, and
  J.~H. Waite, editors, \emph{Enceladus and the {I}cy {M}oons of {S}aturn},
  pages 57--78. University of Arizona Press, 2018.
\newblock ISBN 9780816537075.
\newblock \doi{10.2307/j.ctv65sw2b.12}.

\bibitem[{Hemingway} et~al.(2020){Hemingway}, {Rudolph}, and {Manga}]{Hem2020}
D.~J. {Hemingway}, M.~L. {Rudolph}, and M.~{Manga}.
\newblock {Cascading parallel fractures on Enceladus}.
\newblock \emph{Nat. Astron.}, 4:\penalty0 234--239, 2020.
\newblock \doi{10.1038/s41550-019-0958-x}.

\bibitem[{Hurford} et~al.(2007){Hurford}, {Helfenstein}, {Hoppa}, {Greenberg},
  and {Bills}]{Hur2007}
T.~A. {Hurford}, P.~{Helfenstein}, G.~V. {Hoppa}, R.~{Greenberg}, and B.~G.
  {Bills}.
\newblock {Eruptions arising from tidally controlled periodic openings of rifts
  on {E}nceladus}.
\newblock \emph{Nature}, 447:\penalty0 292--294, 2007.
\newblock \doi{10.1038/nature05821}.

\bibitem[{Kalousov{\'a}} et~al.(2016){Kalousov{\'a}}, {Sou{\v c}ek}, {Tobie},
  {Choblet}, and {{\v C}adek}]{Kal2016}
K.~{Kalousov{\'a}}, O.~{Sou{\v c}ek}, G.~{Tobie}, G.~{Choblet}, and O.~{{\v
  C}adek}.
\newblock {Water generation and transport below {E}uropa's strike-slip faults}.
\newblock \emph{J. Geophys. Res.}, 121:\penalty0 2444--2462, 2016.
\newblock \doi{10.1002/2016JE005188}.

\bibitem[Katz and Worster(2008)]{Kat2008b}
R.~F. Katz and M.~G. Worster.
\newblock Simulation of directional solidification, thermochemical convection,
  and chimney formation in a {Hele-Shaw} cell.
\newblock \emph{J. Comput. Phys.}, 227\penalty0 (23):\penalty0 9823--9840,
  2008.
\newblock \doi{10.1016/j.jcp.2008.06.039}.

\bibitem[Kite and Rubin(2016)]{Kit2016}
E.~S. Kite and A.~M. Rubin.
\newblock Sustained eruptions on {E}nceladus explained by turbulent dissipation
  in tiger stripes.
\newblock \emph{Proc. Natl. Acad. Sci.}, 113\penalty0 (15):\penalty0
  3972--3975, 2016.
\newblock \doi{10.1073/pnas.1520507113}.

\bibitem[Lister(1990)]{Lis1990}
J.~R. Lister.
\newblock Buoyancy-driven fluid fracture: similarity solutions for the
  horizontal and vertical propagation of fluid-filled cracks.
\newblock \emph{J. Fluid Mech.}, 217:\penalty0 213--239, 1990.
\newblock \doi{10.1017/S0022112090000696}.

\bibitem[Nimmo(2004)]{Nim2004b}
F.~Nimmo.
\newblock Stresses generated in cooling viscoelastic ice shells: Application to
  {E}uropa.
\newblock \emph{J. Geophys. Res.}, 109\penalty0 (E12), 2004.
\newblock \doi{10.1029/2004JE002347}.

\bibitem[Nimmo and Gaidos(2002)]{Nim2002}
F.~Nimmo and E.~Gaidos.
\newblock Strike-slip motion and double ridge formation on {E}uropa.
\newblock \emph{J. Geophys. Res.}, 107\penalty0 (E4):\penalty0 5--1--5--8,
  2002.
\newblock \doi{10.1029/2000JE001476}.

\bibitem[Nimmo et~al.(2007)Nimmo, Spencer, Pappalardo, and Mullen]{Nim2007}
F.~Nimmo, J.~R. Spencer, R.~T. Pappalardo, and M.~E. Mullen.
\newblock Shear heating as the origin of the plumes and heat flux on
  {E}nceladus.
\newblock \emph{Nature}, 447\penalty0 (7142):\penalty0 289, 2007.
\newblock \doi{10.1038/nature05783}.

\bibitem[Nimmo et~al.(2011)Nimmo, Bills, and Thomas]{Nim2011}
F.~Nimmo, B.~G. Bills, and P.~C. Thomas.
\newblock Geophysical implications of the long-wavelength topography of the
  {Saturnian} satellites.
\newblock \emph{J. Geophys. Res.}, 116\penalty0 (E11), 2011.
\newblock \doi{10.1029/2011JE003835}.

\bibitem[Nimmo et~al.(2018)Nimmo, Barr, B\v{e}hounkov\'{a}, and
  McKinnon]{Nim2018}
F.~Nimmo, A.~C. Barr, M.~B\v{e}hounkov\'{a}, and W.~B. McKinnon.
\newblock The thermal and orbital evolution of {E}nceladus: Observational
  constraints and models.
\newblock In P.~M. Schenk, R.~N. Clark, C.~J.~A. Howett, A.~J. Verbiscer, and
  J.~H. Waite, editors, \emph{Enceladus and the {I}cy {M}oons of {S}aturn},
  pages 79--94. University of Arizona Press, 2018.
\newblock ISBN 9780816537075.
\newblock \doi{10.2307/j.ctv65sw2b.13}.

\bibitem[Nye(1955)]{Nye1955}
J.~F. Nye.
\newblock Comments on {Dr. Loewe's} letter and notes on crevasses.
\newblock \emph{J. Glaciol.}, 2\penalty0 (17):\penalty0 512--514, 1955.
\newblock \doi{10.3189/S0022143000032652}.

\bibitem[{Olgin} et~al.(2011){Olgin}, {Smith-Konter}, and
  {Pappalardo}]{Olg2011}
J.~G. {Olgin}, B.~R. {Smith-Konter}, and R.~T. {Pappalardo}.
\newblock {Limits of Enceladus's ice shell thickness from tidally driven tiger
  stripe shear failure}.
\newblock \emph{Geophys. Res. Lett.}, 38:\penalty0 L02201, 2011.
\newblock \doi{10.1029/2010GL044950}.

\bibitem[Parkinson et~al.(2020)Parkinson, Martin, Wells, and Katz]{Par2020}
J.~R.~G. Parkinson, D.~F. Martin, A.~J. Wells, and R.~F. Katz.
\newblock Modelling binary alloy solidification with adaptive mesh refinement.
\newblock \emph{J. Comput. Phys.}, 5:\penalty0 100043, 2020.
\newblock \doi{10.1016/j.jcpx.2019.100043}.

\bibitem[Polashenski et~al.(2017)Polashenski, Golden, Perovich, Skyllingstad,
  Arnsten, Stwertka, and Wright]{Pol2017}
C.~Polashenski, K.~M. Golden, D.~K. Perovich, E.~Skyllingstad, Alexandra
  Arnsten, C.~Stwertka, and N.~Wright.
\newblock Percolation blockage: A process that enables melt pond formation on
  first year arctic sea ice.
\newblock \emph{J. Geophys. Res.}, 122\penalty0 (1):\penalty0 413--440, 2017.
\newblock \doi{10.1002/2016JC011994}.

\bibitem[Rudolph et~al.(2022)Rudolph, Manga, Walker, and Rhoden]{Rud2022}
M.~L. Rudolph, M.~Manga, M.~Walker, and A.~R. Rhoden.
\newblock Cooling crusts create concomitant cryovolcanic cracks.
\newblock \emph{Geophys. Res. Lett.}, 49\penalty0 (5):\penalty0 e2021GL094421,
  2022.
\newblock \doi{10.1029/2021GL094421}.

\bibitem[Schulson and Duval(2009)]{Sch2009}
E.~M. Schulson and P.~Duval.
\newblock \emph{Creep and fracture of ice}.
\newblock Cambridge University Press, 2009.

\bibitem[Shibley and Laughlin(2021)]{Shi2021}
N.~C. Shibley and G.~Laughlin.
\newblock Do oceanic convection and clathrate dissociation drive {Europa’s}
  geysers?
\newblock \emph{Planet. Sci. J.}, 2\penalty0 (6):\penalty0 221, 2021.
\newblock \doi{10.1029/2022GL098621}.

\bibitem[Smith(1976)]{Smi1976}
R.~A. Smith.
\newblock The application of fracture mechanics to the problem of crevasse
  penetration.
\newblock \emph{J. Glaciol.}, 17\penalty0 (76):\penalty0 223--228, 1976.
\newblock \doi{10.3198/1976JoG17-76-223-228}.

\bibitem[Smith-Konter and Pappalardo(2008)]{Smi2008}
B.~Smith-Konter and R.~T. Pappalardo.
\newblock Tidally driven stress accumulation and shear failure of {E}nceladus's
  tiger stripes.
\newblock \emph{Icarus}, 198\penalty0 (2):\penalty0 435--451, 2008.
\newblock \doi{10.1016/j.icarus.2008.07.005}.

\bibitem[Spencer and Nimmo(2013)]{Spe2013}
J.~R. Spencer and F.~Nimmo.
\newblock Enceladus: An active ice world in the {Saturn} system.
\newblock \emph{Annu. Rev. Earth Planet. Sci.}, 41\penalty0 (1):\penalty0
  693--717, 2013.
\newblock \doi{10.1146/annurev-earth-050212-124025}.

\bibitem[Stevenson(1996)]{Ste1996}
D.~J. Stevenson.
\newblock Heterogeneous tidal deformation and geysers on {Europa}.
\newblock In \emph{Europa Ocean Conference}, number~5 in Capistrano conference,
  pages 69--70, San Juan Capistrano, Calif., 1996. San Juan Capistrano Res.
  Inst.

\bibitem[Tait and Jaupart(1992)]{Tai1992}
S.~Tait and C.~Jaupart.
\newblock Compositional convection in a reactive crystalline mush and melt
  differentiation.
\newblock \emph{J. Geophys. Res.}, 97\penalty0 (B5):\penalty0 6735--6756, 1992.
\newblock \doi{10.1029/92JB00016}.

\bibitem[Timco and Frederking(1996)]{Tim1996}
G.~W. Timco and R.~M.~W. Frederking.
\newblock A review of sea ice density.
\newblock \emph{Cold Reg. Sci. Technol.}, 24\penalty0 (1):\penalty0 1--6, 1996.
\newblock ISSN 0165-232X.
\newblock \doi{10.1016/0165-232X(95)00007-X}.

\bibitem[Walker et~al.(2021)Walker, Bassis, and Schmidt]{Wal2021}
C.~C. Walker, J.~N. Bassis, and B.~E. Schmidt.
\newblock Propagation of vertical fractures through planetary ice shells: The
  role of basal fractures at the ice–ocean interface and proximal cracks.
\newblock \emph{Planet. Sci. J.}, 2\penalty0 (4):\penalty0 135, 2021.
\newblock \doi{10.3847/PSJ/ac01ee}.

\bibitem[Wells et~al.(2019)Wells, Hitchen, and Parkinson]{Wel2019}
A.~J. Wells, J.~R. Hitchen, and J.~R.~G. Parkinson.
\newblock Mushy-layer growth and convection, with application to sea ice.
\newblock \emph{Philos. Trans. R. Soc. London, Ser. A}, 377\penalty0
  (2146):\penalty0 20180165, 2019.
\newblock \doi{10.1098/rsta.2018.0165}.

\end{thebibliography}
\end{document}


\title{\vspace{-1.5cm}Supplement: ``A mushy source for the geysers of Enceladus''}
\author[1]{Colin R. Meyer\thanks{colin.r.meyer@dartmouth.edu}}
\author[1]{Jacob J. Buffo}
\author[2]{Francis Nimmo}
\author[3]{Andrew J. Wells}
\author[4]{Samuel Boury}
\author[1]{Tara C. Tomlinson}
\author[3]{Jamie R. G. Parkinson}
\author[5]{Geoffrey M. Vasil}
\affil[1]{Thayer School of Engineering, Dartmouth College, Hanover, NH 03755}
\affil[2]{Department of Earth \& Planetary Sciences, University of California, Santa Cruz, CA 95064 USA}
\affil[3]{Atmospheric, Oceanic \& Planetary Physics, Department of Physics, University of Oxford, OX1 3PU, UK}
\affil[4]{Courant Institute of Mathematical Sciences, New York University, New York, NY 10012}
\affil[5]{School of Mathematics, University of Edinburgh, EH9 3FD, UK}
\maketitle

\section{Fracture depths}
It is unclear how easy it is to produce a fracture spanning the entire ice shell of Enceladus. Due to the limited  tidal stresses ($\sim100$~kPa), when compared to the overburden pressure ($\sim600$ kPa at the base of a 6 km ice shell), it is hard for near-surface fractures to extend downwards more than 1--2~km \citep{Smi2008,Olg2011,Wal2021}. This is less than the estimated minimum shell thickness of 6~km at the south pole \citep{Hem2018}.  Fractures extending upwards from the base of the shell are less affected by overburden pressure because they would be water-filled \citep{Cra1988,Shi2021}. However, ice at the base of the shell is more ductile and thus harder to fracture than ice at the surface \citep{Sch2009}. 

The current paradigm for the source of geyser material is that fractures extend down to the subsurface ocean \citep{Hur2007}. To get an idea for the depth of the tiger stripe fractures, we balance the extensional stress required to open a fracture with the overburden ice pressure acting to close it (i.e. the zero-stress \citet{Nye1955} criterion for closely spaced glacier crevasses). Estimates for the extensional stress are on the order of $\tau\sim1$ bar \citep{Nim2007,Nim2011}. Gravity on Enceladus is $g=0.113$ m s$^{-2}$ \citep{Han2006,Han2008} and the salty ice density is about $\rho_{\textrm{ice}}=940$ kg m$^{-3}$ \citep{Tim1996}.  The fracture close-off depth $h_{\textrm{top}}$ is given as
\begin{equation}
h_{\textrm{top}} \sim \frac{\tau}{\rho_{\textrm{ice}} g} \sim 1~\mathrm{km},
\label{eqn:Nye55t}
\end{equation}
which is an estimate for the likely depth of surface fractures and agrees with more sophisticated calculations \citep{Smi1976,Bro1999,Smi2008,Olg2011}. The fact that the surface fractures are only likely to propagate to $\sim1~\mathrm{km}$ depth is important because $\sim1~\mathrm{km}$ is shallower than the minimum predicted shell thickness of $\sim6-30~\mathrm{km}$ \citep{Hem2018}, which casts doubt on the idea that the fractures penetrate from the surface down to the ocean. Moreover, it would be a considerable feat for a fracture to extend up from the ocean (i.e. a basal crevasse) to within $\sim1~\mathrm{km}$ of the surface, based on the extensional stress just mentioned. A water-filled basal crevasse does lead to a larger close-off height $h_{\textrm{bottom}}$, given by 
\begin{equation}
h_{\textrm{bottom}} \sim \frac{\tau}{\left(\rho_{\textrm{water}}-\rho_{\textrm{ice}}\right) g} \sim 10\mbox{ km},
\label{eqn:Nye55b}
\end{equation}
where we use a value of seawater density, $\rho_{\textrm{water}}= 1030$ kg m$^{-3}$. This is sufficient to fracture the shell if (i) the fracture emanates up from the bottom and (ii) if the shell is on the lower end of thickness range \citep[i.e. $\sim 6-30~\mathrm{km}$][]{Hem2018}. However, the lower shell is likely ductile and less prone to fracturing \citep{Nim2004b,Sch2009}. On the other hand, there are additional sources of stress available to open cracks which likely exceed the tidal stresses. These include the effects of shell thickening \citep{Rud2022} and flexural stresses from plume loading \citep{Hem2020}. Thus, opening of through-going fractures appears possible but not certain. At the same time, the dike mechanism discussed in the main text could allow for the initiation of downward-propagating through-going fractures that could be maintained by turbulent water flow and shear heating. We discuss the dike mechanism further in section \ref{sec:dike}.

\section{Analytical temperature solution}
Here we demonstrate that the leading order thermal structure can be understood from steady state conduction, with reactive flow causing subsequent modifications to the porosity structure, composition and shape of the inclusion. We consider a fracture of depth $h$ in a shell of thickness $H$ with a shear heat flux of $\mu \rho_i g h u$. So long as the latent heat sink of partial melting and heat transport by interstitial flow are relatively small, then the steady state temperature $T$ around the fracture is given by the solution to Laplace's equation, i.e.
\begin{equation}
\nabla^2 T = 0,\label{eqn:lapT}
\end{equation}
where we treat the thermal conductivity as temperature independent to facilitate analytical progress. 
Equation \eqref{eqn:lapT} is subject to the boundary conditions
\begin{eqnarray}
-k\pd{T}{x} &=& 0~~~\mbox{on}~~~ x=0~(z<h)~~~\mbox{and}~~~x=L,\\
-k\pd{T}{x} &=& \mu \rho_i g h u~~~\mbox{on}~~~ x=0~(z\geq h),\\
-k\pd{T}{z} &=& r_1 + r_2 (T-T_s)~~~\mbox{on}~~~z=H,\\
T&=&T_e~~~\mbox{on}~~~z=0.
\end{eqnarray}
The surface condition is a linearized radiative condition, with respect to the background temperature $T_s$ of space near Enceladus (see table \ref{prms}), rather than the full black-body radiation condition as used by \citet{Abr2009}. This choice for the surface boundary condition allows for an analytical solution and is the condition used in SOFTBALL. To derive the linearized condition, we start from the surface energy balance where conductive heat fluxes from the ice balance net black-body radiation as
\begin{equation}
-k \pd{T}{z} = \epsilon \sigma \left( T^4 - T_s^4\right),
\end{equation}
where $\epsilon$ is the emissivity and $\sigma$ is the Stefan-Boltzmann constant (values in table \ref{prms}). Expanding the temperature as a small deviation from the space temperature, i.e. $T=T_s+\tilde{T}$, where $\tilde{T}\ll T_s$, then we can use the binomial expansion and discard terms of second order and higher, to find that 
\begin{equation}
-k \pd{T}{z} = 4 \epsilon \sigma T_s^3 (T-T_s) ,
\end{equation}
which is in the form of the boundary condition above, with $r_1=0$ and $r_2 = 4\epsilon \sigma T_s^3$.

We nondimensionalized the problem, by taking 
\[T=T_e+\Delta T \theta,~~~x = H \tilde{x},~~~z= H \tilde{z},\]
which yields
\begin{equation}
\tilde{\nabla}^2 \theta = 0,\label{eqn:laptheta}
\end{equation}
subject to the boundary conditions
\begin{eqnarray}
-\pd{\theta}{\tilde{x}} &=& 0~~~\mbox{on}~~~ \tilde{x}=0~(\tilde{z}<\delta )~~~\mbox{and}~~~\tilde{x}= \lambda,\\
-\pd{\theta}{\tilde{x}} &=& F~~~\mbox{on}~~~ \tilde{x}=0~(\tilde{z}\geq \delta),\\
-\pd{\theta}{\tilde{z}} &=&   G  + b\left(\theta-\theta_s\right) ~~~\mbox{on}~~~\tilde{z}=1,\\
\theta&=&0~~~\mbox{on}~~~\tilde{z}=0.
\end{eqnarray}
Here we define the parameters as
\[\delta=\frac{h}{H},~~~\lambda = \frac{L}{H},~~~F = \frac{\mu \rho_i g h H u}{k_o \Delta T},~~~G = \frac{Hr_1}{k_o\Delta T},~~~b = \frac{H r_2}{k_o},~~~\theta_s = \frac{T_s-T_e}{\Delta T}.\]
Due to convergence issues in the SOFTBALL simulations, we could not use large values for $b$, and so the value quoted in table \ref{prms} is artificially lower than what we would calculate from the parameters. The main effect of this difference is on the surface temperature profile and not the size, shape, or liquid volume of the mushy zone. Asymptotically, our analytical solutions shows that large values of $b$ pin the surface temperature to the space temperature, except very near the crack, where there is a narrow boundary layer region of rapid temperature change. The boundary layer becomes thinner as $b$ increases. Additionally, since $G=0$ here, we disregard it in subsequent calculations.

To simplify the problem in the vertical direction, we write the total solution $\theta$ as a sum of a particular solution $\theta_I$ to the inhomogeneous problem and the homogenous solution $\theta_H$, i.e.
\[\theta = \theta_I(\tilde{z}) + \theta_{H}(\tilde{x},\tilde{z}).\]
Solving first for the inhomogeneous part, we have that 
\[\pd{^2\theta_I}{\tilde{x}^2}+\pd{^2\theta_I}{\tilde{z}^2} = 0 \longrightarrow \pd{^2\theta_I}{\tilde{z}^2} = 0,\]
so 
\[\theta_I = A \tilde{z} + B.\]
Using the boundary conditions, we find that
\[B = 0\ands A = \frac{b \theta_s}{1+b}.\]
For large values of $b$, the slope $A$ is approximately $\theta_s$, which is consistent with the asymptotic description above, that the surface temperature mostly stays close to the space temperature. 

\begin{figure}
\includegraphics[width=\linewidth]{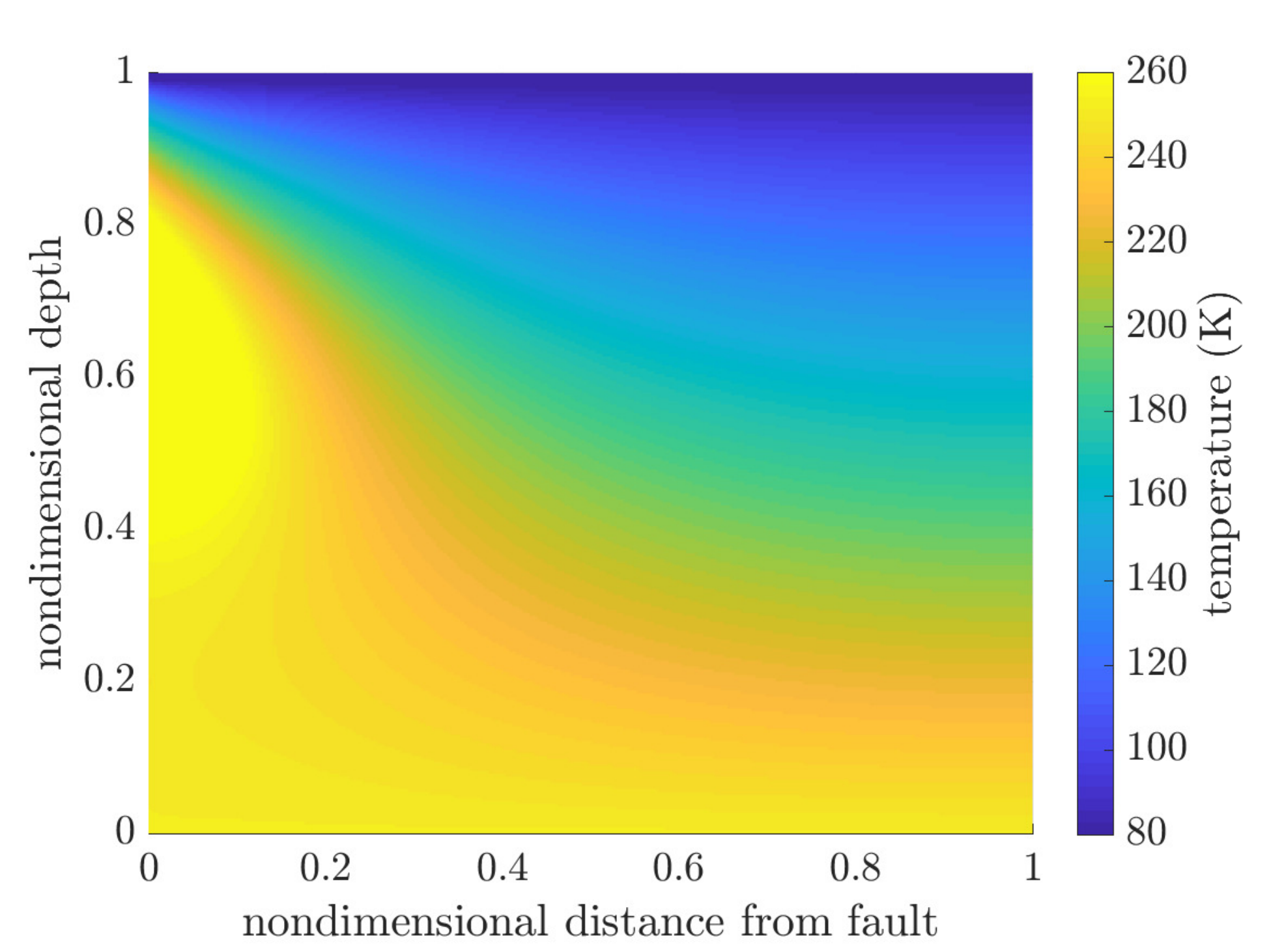}
\caption{Analytical solution for the temperature around a heated crack with $F=260$.}
\label{fig:stevenson_closer}
\end{figure}

Now we can examine the homogeneous problem using separation of variables, i.e. 
\[\theta_H(\tilde{x},\tilde{z}) = f(\tilde{x})g(\tilde{z}),~~~\mbox{which implies}~~~f'' g + f g'' = 0.\]
Dividing through and choosing the separation constant $\alpha^2$, we find the two ODEs
\[f'' =\alpha^2 f\ands g'' = -\alpha^2 g.\]
Solving the vertical equation first, we have that 
\[g = C \sin(\alpha \tilde{z})+D\cos(\alpha \tilde{z}).\]
With $g=0$ at $\tilde{z}=0$, we know that $D=0$. The homogeneous surface boundary condition is
\[ - g' = b g\spword{at}\tilde{z}=1,\]
which results in the transcendental eigenvalue condition 
\begin{equation}
- \alpha \cos(\alpha) = b  \sin(\alpha).\label{eqn:tec}
\end{equation}
This yields a full set of distinct eigenvalues $\alpha_n$ for $n=1,2,3,...$, which can be computed numerically. In the horizontal direction, we have that 
\[f = D e^{\alpha \tilde{x}} + E e^{-\alpha \tilde{x}}.\]
Using the no heat flux boundary condition at $\tilde{x} = \lambda$, we have that 
\[-f' = 0\spword{at}\tilde{x}=\lambda \spword{implies} E = D e^{2\alpha \lambda }.\]
Thus, the full solution with unknown coefficients is given by 
\begin{equation}
\theta = \frac{b \theta_s}{1+b} \tilde{z} + \sum_{n=1}^{\infty}{c_n \left[ e^{\alpha_n \tilde{x}} + e^{2\alpha_n \lambda } e^{-\alpha_n \tilde{x}} \right] \sin(\alpha_n \tilde{z})}.
\end{equation}
We find the values for $c_n$ by using orthogonality and integrating over the heat flux condition at $\tilde{x} = 0$, which gives
\begin{equation}
\int_{\delta}^{1}{F\sin(\alpha_m\tilde{z})~d\tilde{z}} = \sum_{n=1}^{\infty}{c_n \alpha_n \left[1 - e^{2\alpha_n \lambda }  \right]  \int_{0}^{1}{\sin(\alpha_n\tilde{z})\sin(\alpha_m\tilde{z})~d\tilde{z}}}.
\end{equation}
The integral on right is zero for $\alpha_n \ne \alpha_m$, using the transcendental eigenvalue condition from equation \eqref{eqn:tec} implying orthogonality of the eigenfunctions $\sin(\alpha_n\hat{z})$. When $\alpha_n = \alpha_m$, we have that
\[\int_{0}^{1}{\sin(\alpha_m\tilde{z})\sin(\alpha_m\tilde{z})~d\tilde{z}}  = \frac{1}{2} - \frac{\sin(2\alpha_m)}{4\alpha_m}.\]
We evaluate the integral on the left as 
\[\int_{\delta}^{1}{F\sin(\alpha_m \tilde{z})~d\tilde{z}} = -\frac{F}{\alpha_m} \left[\cos(\alpha_m) - \cos(\alpha_m \delta)\right].\]
Putting it all together gives, 
\begin{equation}
c_n = -\frac{4 F }{\left[2\alpha_n^2 - \alpha_n\sin(2\alpha_n)\right]} \frac{\left[\cos(\alpha_n) - \cos(\alpha_n\delta)\right]}{1 - e^{2\alpha_n \lambda }}.
\end{equation}
We show the analytical solution in figure \ref{fig:stevenson_closer} and find that a mushy zone will form along the crack face in regions with $T>T_e$. These results are similar to what we find in the main text as well as \citet{Hamm2020}, \citet{Kal2016}, and \citet{Ste1996}. This analytical solution also allows us to predict the size of the mushy zone, by looking for the contour where $T=T_e$. 
 
\begin{figure}
\includegraphics[width=0.5\linewidth]{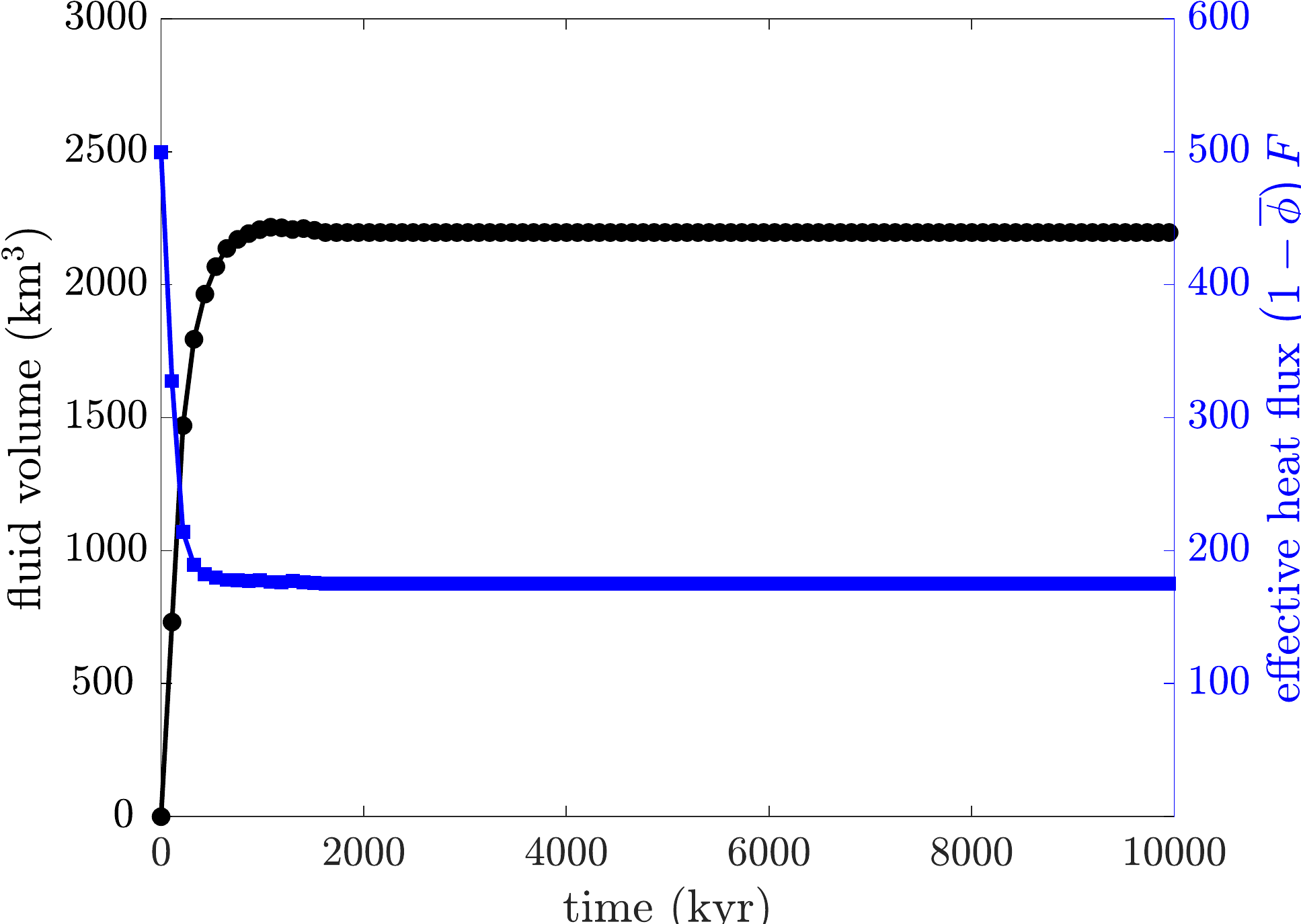}
\includegraphics[width=0.47\linewidth]{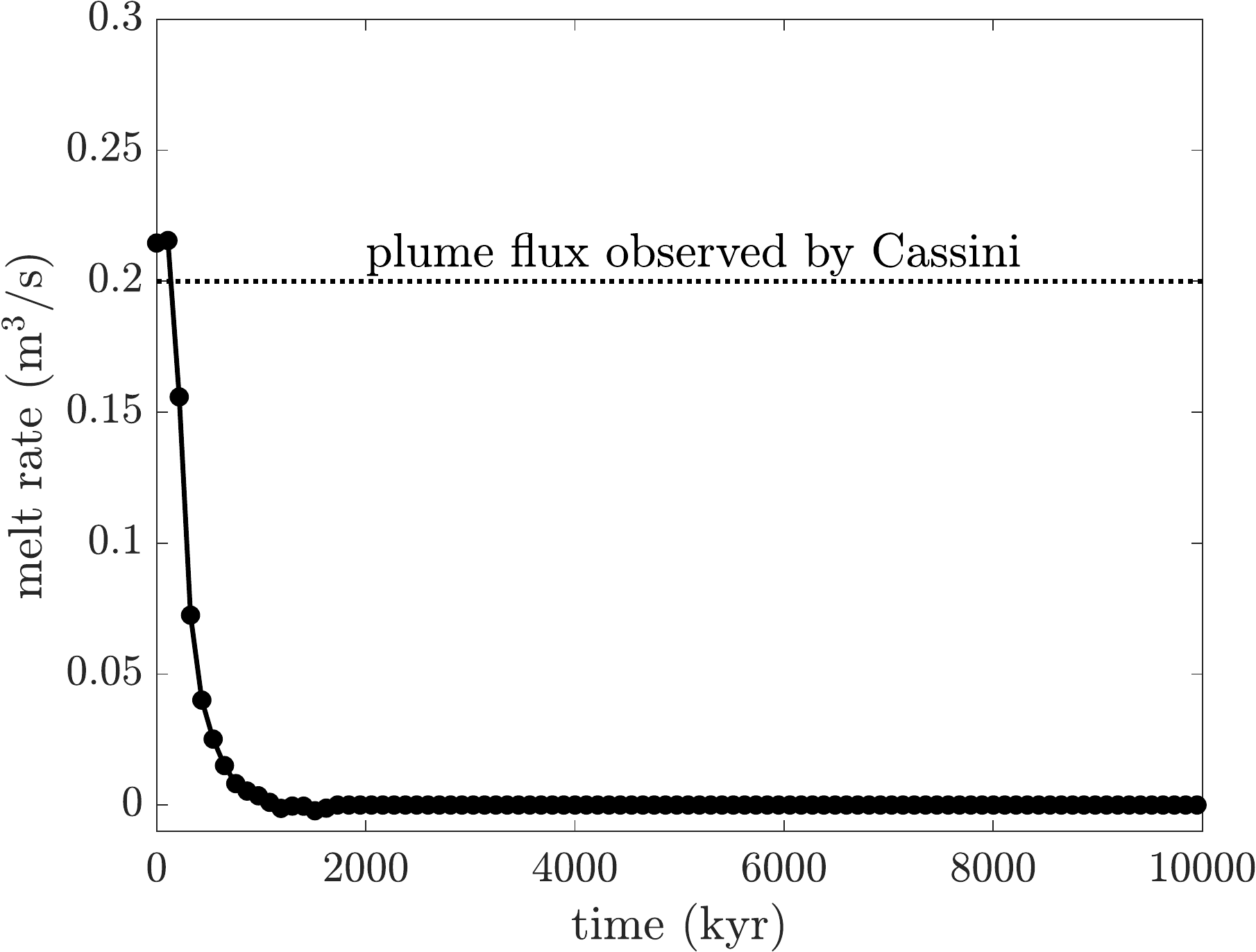}
\caption{Liquid volume in the mushy zone for $F=500$:  (a) evolution to a steady state and average effective heat flux as well as (b) rate of liquid volume production, which can support the entire flux observed by the Cassini spacecraft from a single tiger stripe fracture.}
\label{fig:volume}
\end{figure}

\section{Mushy zone physics}
To analyze the effect of shear heating on a fracture in an icy shell, we use SOFTBALL (SOlidification, Flow and Thermodynamics in Binary ALLoys), a mushy layer code with adaptive mesh refinement. The github repository for this open-source code is: https://github.com/jrgparkinson/mushy-layer (accessed 8 April 2023). This state-of-the-art code solves the equations of mushy layer theory, as described below, in two-dimensional geometries \citep{Wel2019,Par2020}. The code uses a finite volume discretization and adaptive time steps. In our simulations we use a spatial scale of $N_x = 512$ grid points in the horizontal direction and $N_z = 256$ in the vertical direction. This gives a resolution of $\sim40$ meters in both the horizontal and vertical directions. The time steps adjust and are on the order of $\sim30$ years. The simulations take on the order of a day to run on the linux system (2-16 cores) at Dartmouth College. 

Conservation of mass in a mushy layer dictates that 
\begin{equation}
\pd{\left(\phi\rho\right)}{t} = M_f ~~~\mbox{and}~~~\pd{\left[\left(1-\phi\right)\rho\right]}{t} + \bs{\nabla}\cdot\bs{q}= -M_f,\label{eqn:masscons}
\end{equation}
where the solid and liquid densities $\rho$ are approximated as equal \citep[using an extended form of the Boussinesq approximation; ][]{Par2020}, the solid fraction is $\phi$, the fluid flux relative to the solid is $\boldsymbol{q}$, and the rate of mass phase change per volume is $M_f$. The sum of equations \eqref{eqn:masscons} yields $\bs{\nabla}\cdot \bs{q}=0$, i.e. a solenoidal fluid flow field. Here we assume that the porous solid matrix is stationary. Fluid flux is a function of pressure gradients and gravity, which can be described by Darcy's law, as 
\begin{equation}
\boldsymbol{q}= -\frac{\Pi}{\eta}\left( \nabla p + \rho_{b}g\boldsymbol{z} \right),\label{eqn:Darcy}
\end{equation}
where $\eta$ is the liquid viscosity, $\rho_b$ is the density of liquid brine (which is a function of the brine concentration and temperature), $\boldsymbol{z}$ is the vertical unit vector, and $\Pi$ is the solid-fraction-dependent permeability \citep{Tai1992,Kat2008b}, given by
\begin{equation}
\Pi = \left[ \frac{12}{d^2} +\frac{1}{K_0\phi^3} \right]^{-1}.
\end{equation}
Following \citet{Kat2008b}, SOFTBALL uses a permeability regularization for pure liquid regions (i.e. as $\phi\rightarrow1$) such that there is a background permeability and Darcy's law is solved everywhere, rather solving Navier-Stokes in the pure fluid region \citep[see][]{Par2020,Buf2021b}. The ratio of the background permeability $d^2/12$ to the permeability prefactor $K_0$ is the reluctance $\mathtt{R} = d^2/(12K_0)$. 

Conservation of energy in the mushy layer is given by 
\begin{equation}
\overline{\rho c_p} \pd{T}{t} + \rho_b c_b \bs{q}\cdot\bs{\nabla}T = \bs{\nabla}\cdot\left(\overline{k}\bs{\nabla}T\right)+\mathscr{L}M_f,\label{eqn:coe}
\end{equation}
where $\overline{\rho c_p}$ and $\overline{k}$ are the solid-fraction-weighted functions of ice and brine properties for heat capacity and thermal conductivity, respectively. The latent heat is $\mathscr{L}$. We write a conservation law for the concentration of brine, i.e. 
\begin{equation}
\left(1-\phi\right) \pd{C}{t} + \boldsymbol{q}\cdot\boldsymbol{\nabla}C = \boldsymbol{\nabla}\cdot\left(\overline{D}\boldsymbol{\nabla}C\right)+\frac{\left(C-C_s\right)}{\rho}M_f,\label{eqn:cob}
\end{equation}
where $\overline{D}$ is the phase-weighted salt diffusivity arising from transport through the interstitial liquid. 

The evolution of temperature and concentration are coupled in the mushy region, as shown in figure 1 in the main text, where the ice and liquid phases are at an equilibrium temperature given by the liquidus. This imposes a constraint on the evolution equations (\ref{eqn:coe}) and (\ref{eqn:cob}) in mushy regions, where the concentration and temperature are tied together by the liquidus relationship $T=T_L(C)$. In idealized form, the liquidus relationship is given as
\begin{equation}
T_L = T_m - m C,\label{eqn:liquidus}
\end{equation}
where $T_m$ is the pure substance melting temperature and $m$ is the liquidus slope. This constraint no longer applies when $T<T_e$, where $M_f=0$, $\bs{q}=\bs{0}$, $\phi=1$, \eqref{eqn:coe} becomes the heat equation for thermal conduction and salt conservation \eqref{eqn:cob} implies constant bulk salinity.

\section{Liquid volume and rates of production}
In the main text, we show steady state temperature and porosity fields. We treat the rate of shear heating as constant, given that the thermal diffusion timescale is much longer than the tidal cycle \citep{Nim2002,Nim2007}. Here, in figure \ref{fig:volume}(left), we show the approach to steady state for the liquid volume in the mushy zone. Initially, there is no mushy zone and therefore no liquid volume. Then a mushy zone forms and the liquid volume increases rapidly. Due to the temperature field reaching a steady state and the porosity-weakening of the fault friction saturates causing the heat flux $F(1-\overline{\phi})$ to saturate in figure \ref{fig:volume}(left). In all simulations, the local value of $\phi$ is used to calculate the shear heating rate. Here we show the average value over fracture as an aggregate measure of the effect of reducing the shear heating flux. From the curve of liquid volume with time, we take the derivative to find the rate of liquid production, as shown in figure \ref{fig:volume}(right). This shows the same three regimes: no initial liquid volume, growth of a mushy zone, and approach to steady state.

\begin{figure}
    \centering
    \begin{overpic}[width=0.32\linewidth]{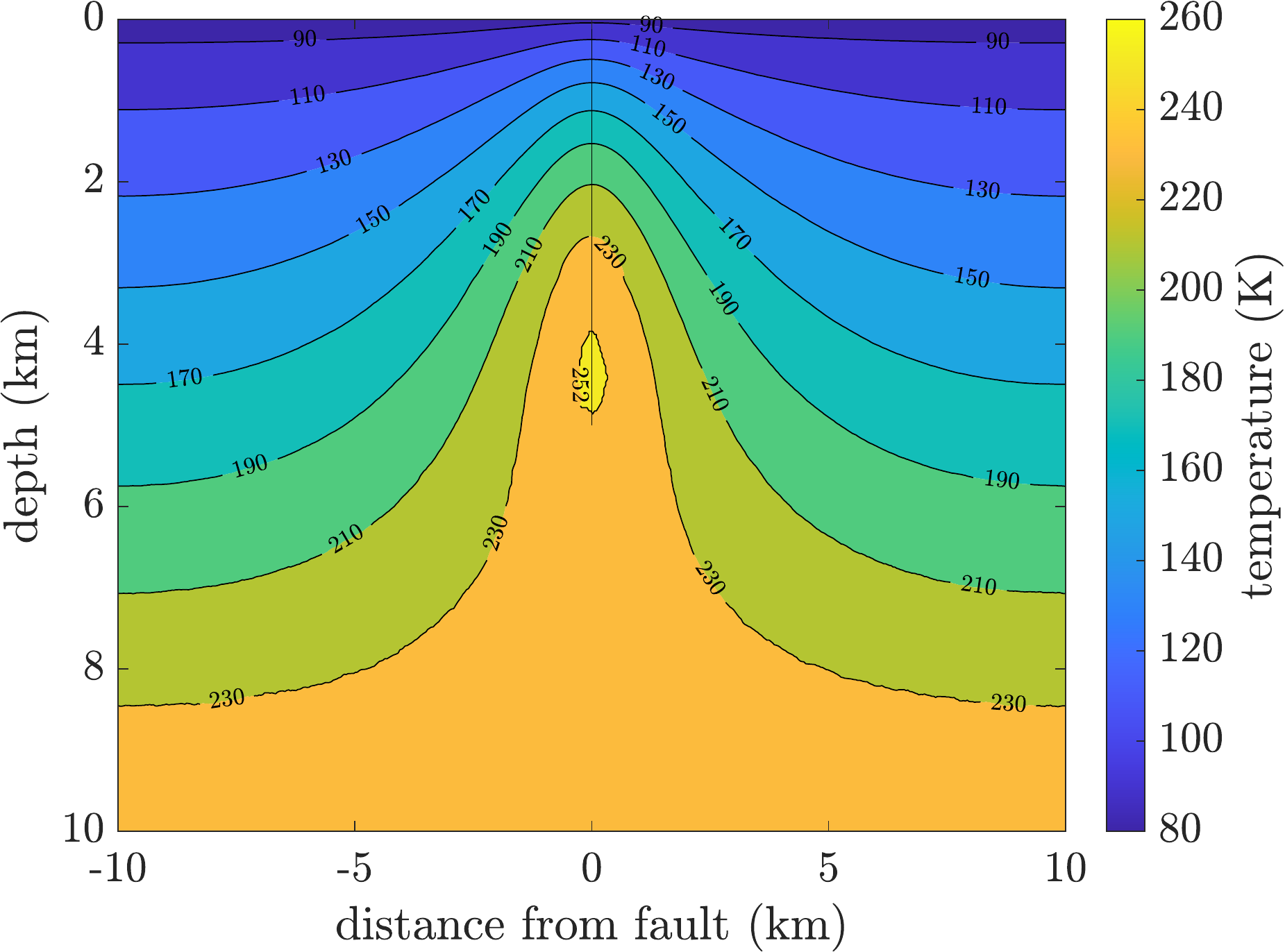}
    \put(33,75){$F = 180$}\end{overpic}
    \begin{overpic}[width=0.32\linewidth]{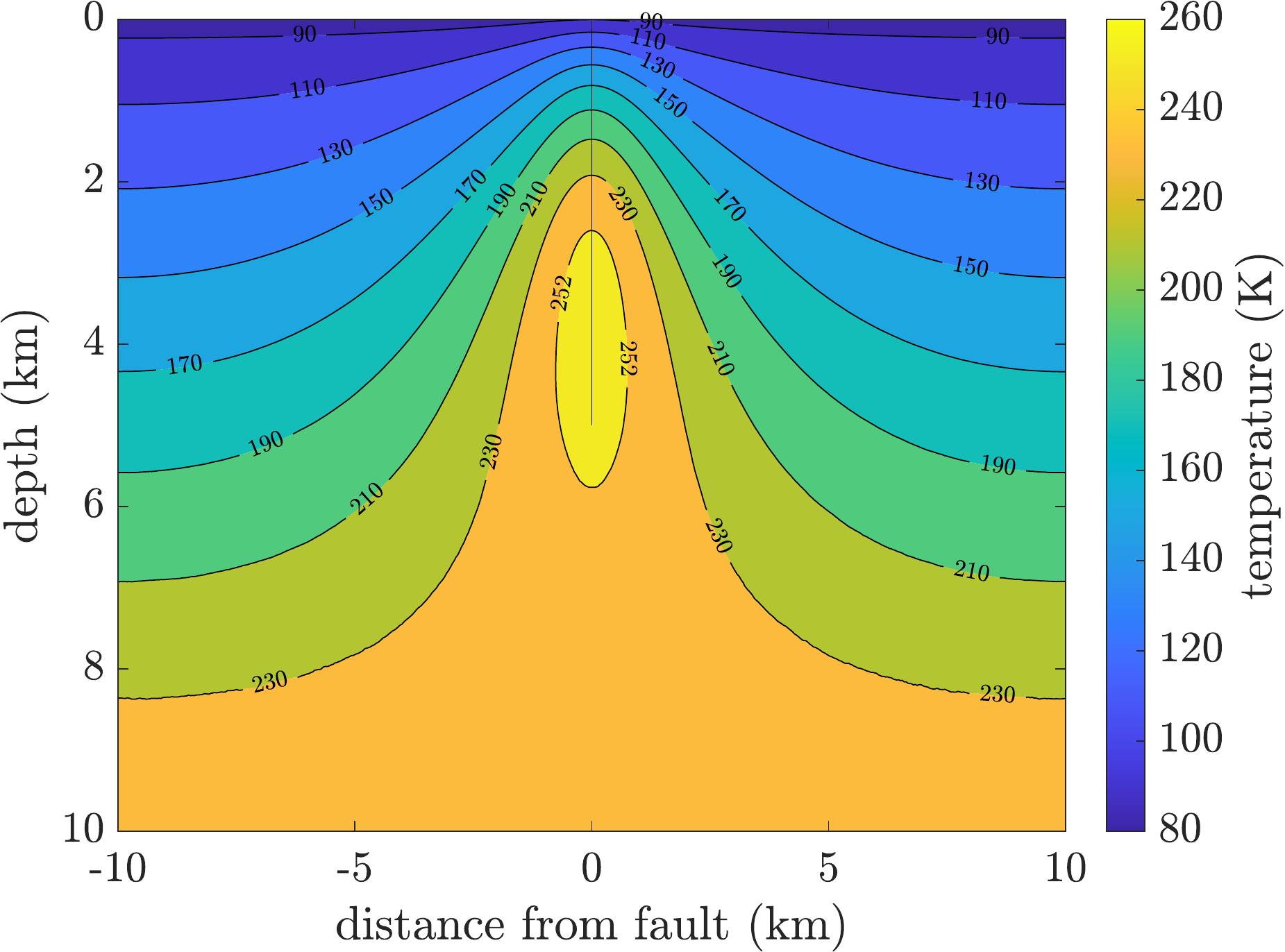}
    \put(33,75){$F = 240$}\end{overpic}
    \begin{overpic}[width=0.32\linewidth]{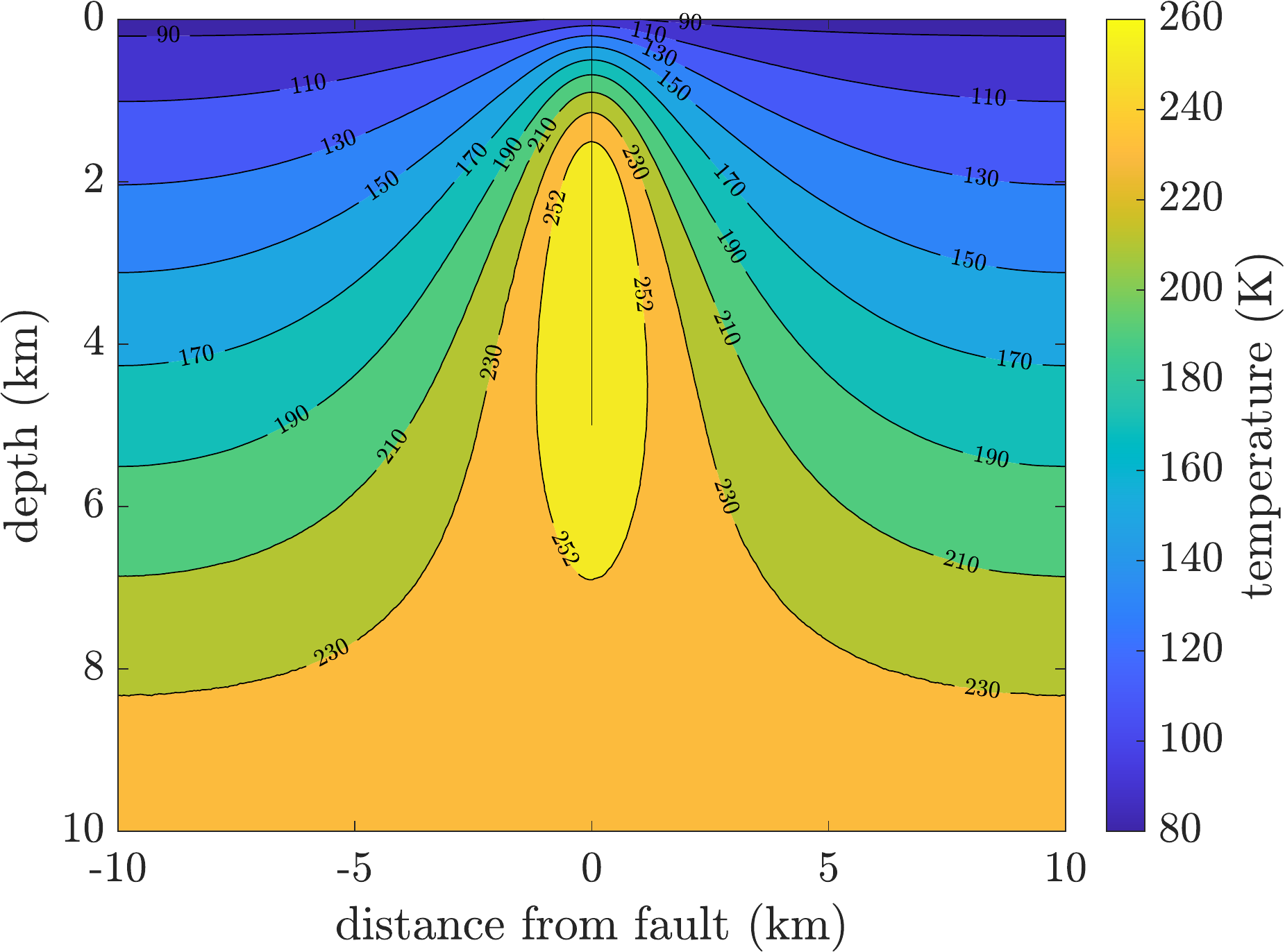}
    \put(33,75){$F = 420$}\end{overpic}
    \begin{overpic}[width=0.32\linewidth]{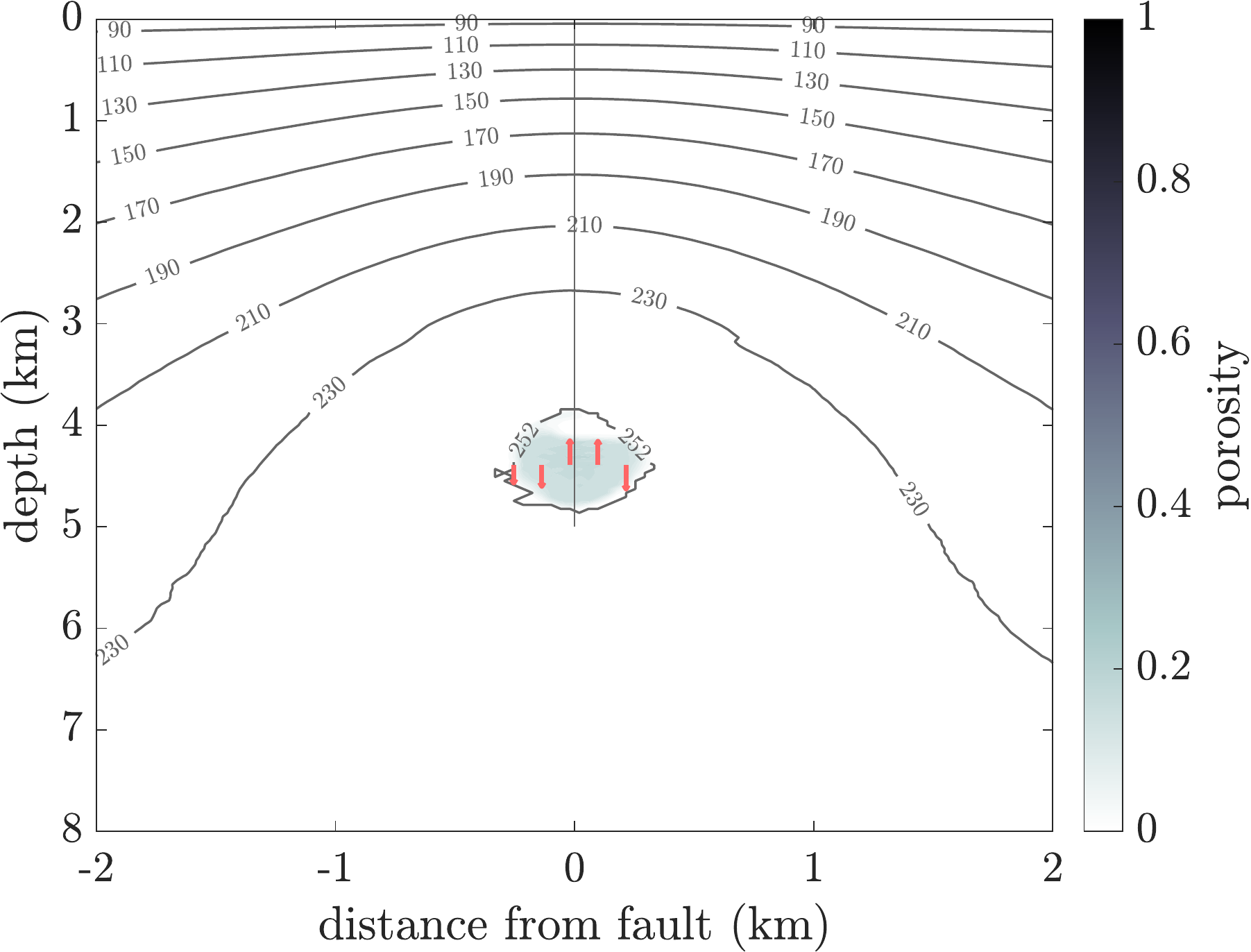}
    \end{overpic}
    \begin{overpic}[width=0.32\linewidth]{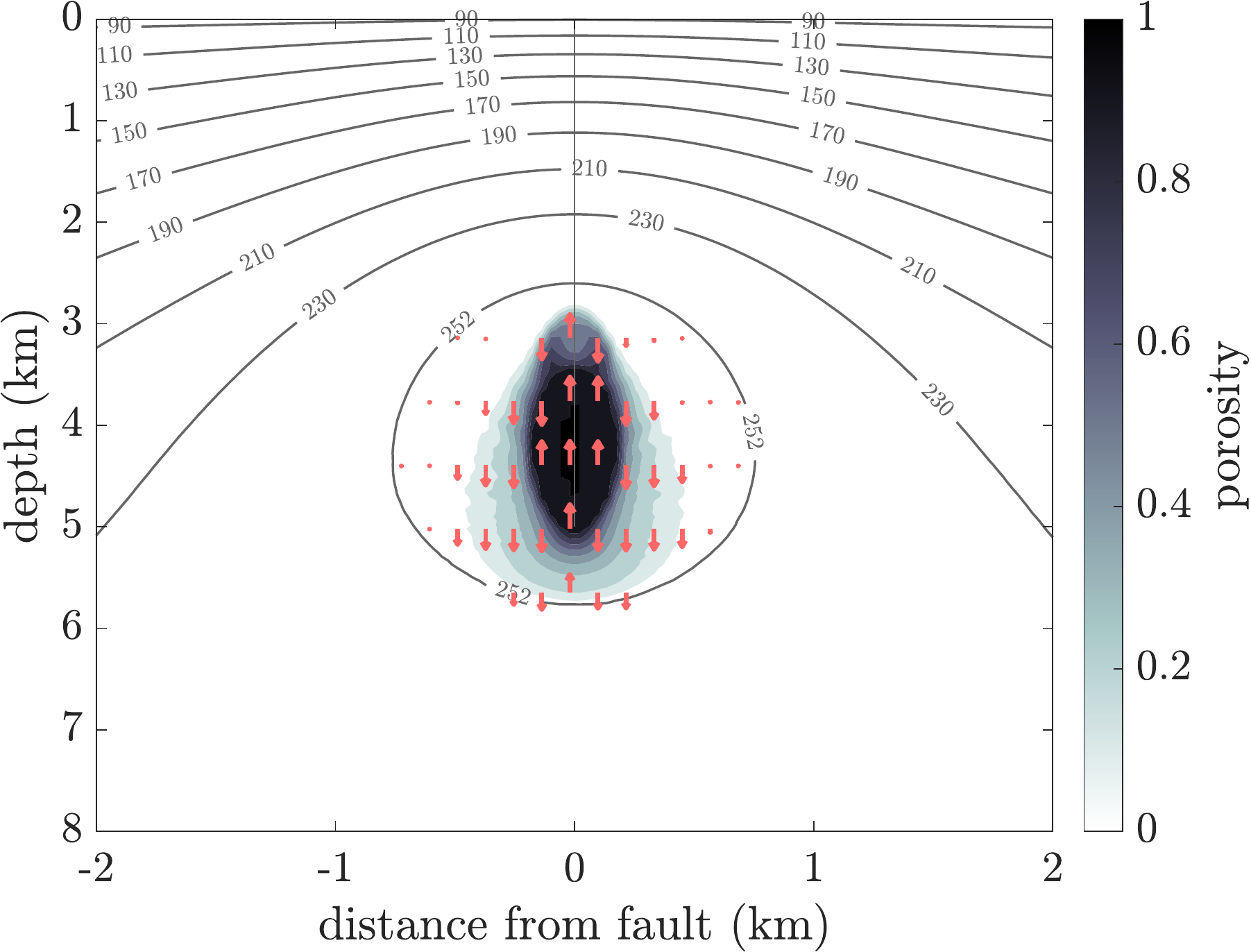}
    \end{overpic}
    \begin{overpic}[width=0.32\linewidth]{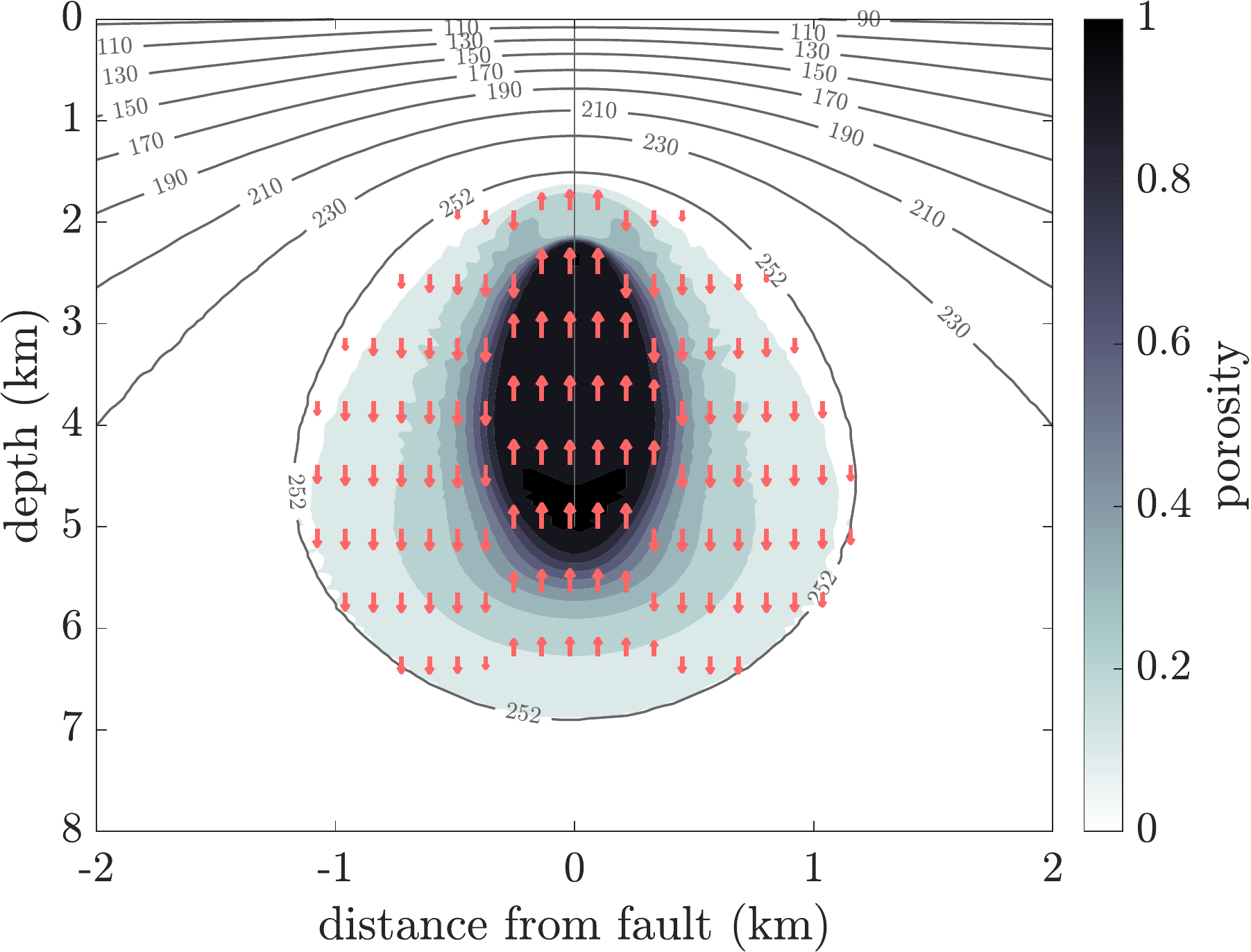}
    \end{overpic}
    \caption{Temperature and porosity plots for three different values of the nondimensional heat flux: (left) $F=180$, (middle) $F=240$, and (right) $F=420$. Note that porosity is shown with different axis than the temperature plots. Similar to the main text, we show vertical velocity arrows with logarithmic magnitude with the largest arrows representing about 1 mm/year.}
    \label{fig:3tempandporosity}
\end{figure}

The observed geyser ejection rate for all four tiger stripes is $\sim$0.2 m$^{3}$ s$^{-1}$ \citep[][]{Han2006,Han2008}.  Following a step change in heating and for the range of heat fluxes we used in these simulations, we find that the maximum rate of liquid volume production is greater than $0.2$ m$^{3}$ s$^{-1}$ ($F=500$), which is sufficient to explain the geyser flux from all of the tiger stripe fractures. Figure \ref{fig:volume} shows that the melt rate is equal to the observed geyser ejection rate initially and stays at a comparable magnitude for $\sim$200 thousand years. At later times, the melt rate reduces and melt volume saturates because a region of large porosity develops and weakens the frictional heating on the crack. However, venting of liquid from the crack will limit porosity growth, and enable sustained heating and melting rates.

\begin{figure}[!ht]
    \centering
    \includegraphics[width=0.49\linewidth]{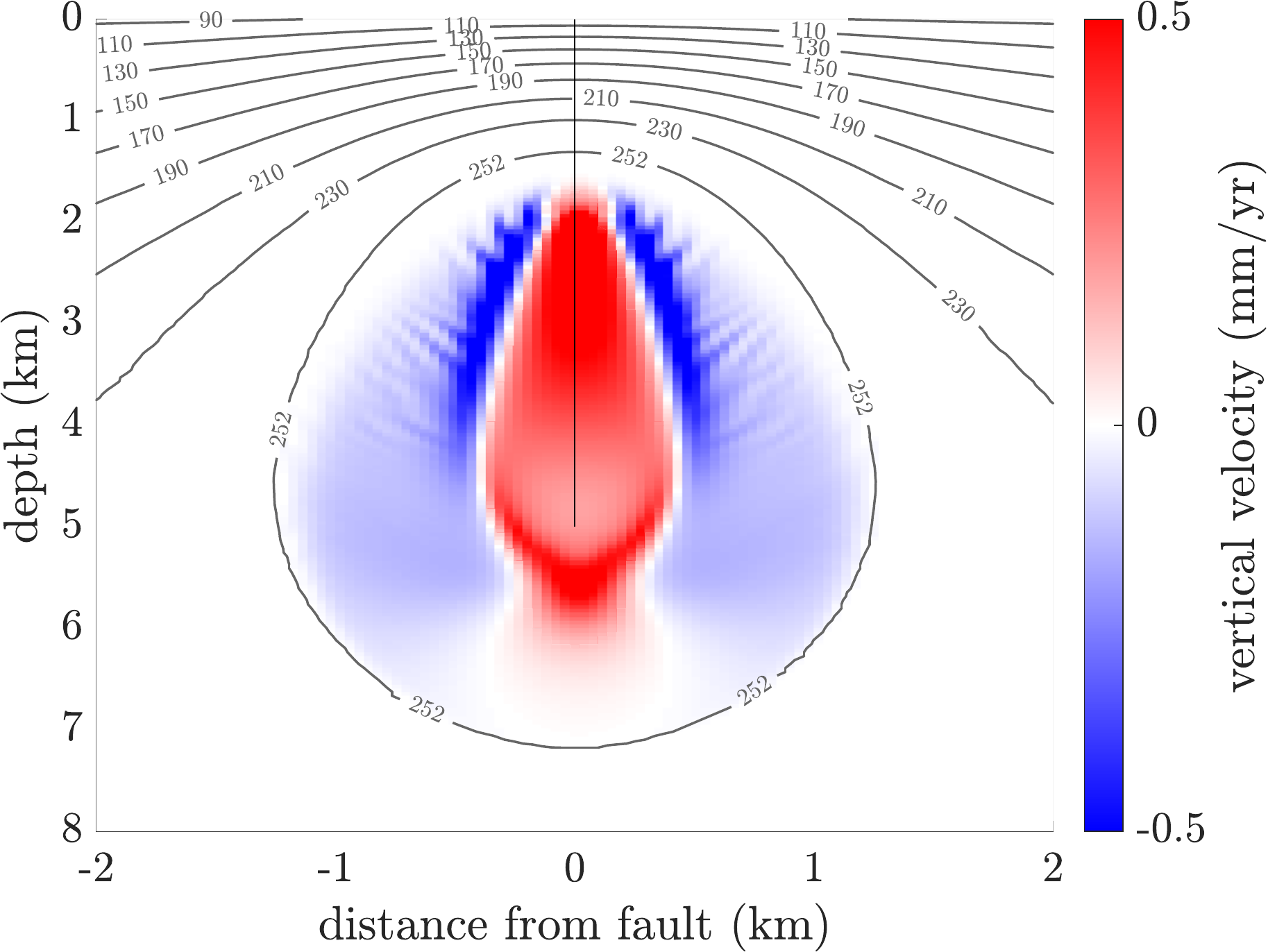}
    \includegraphics[width=0.49\linewidth]{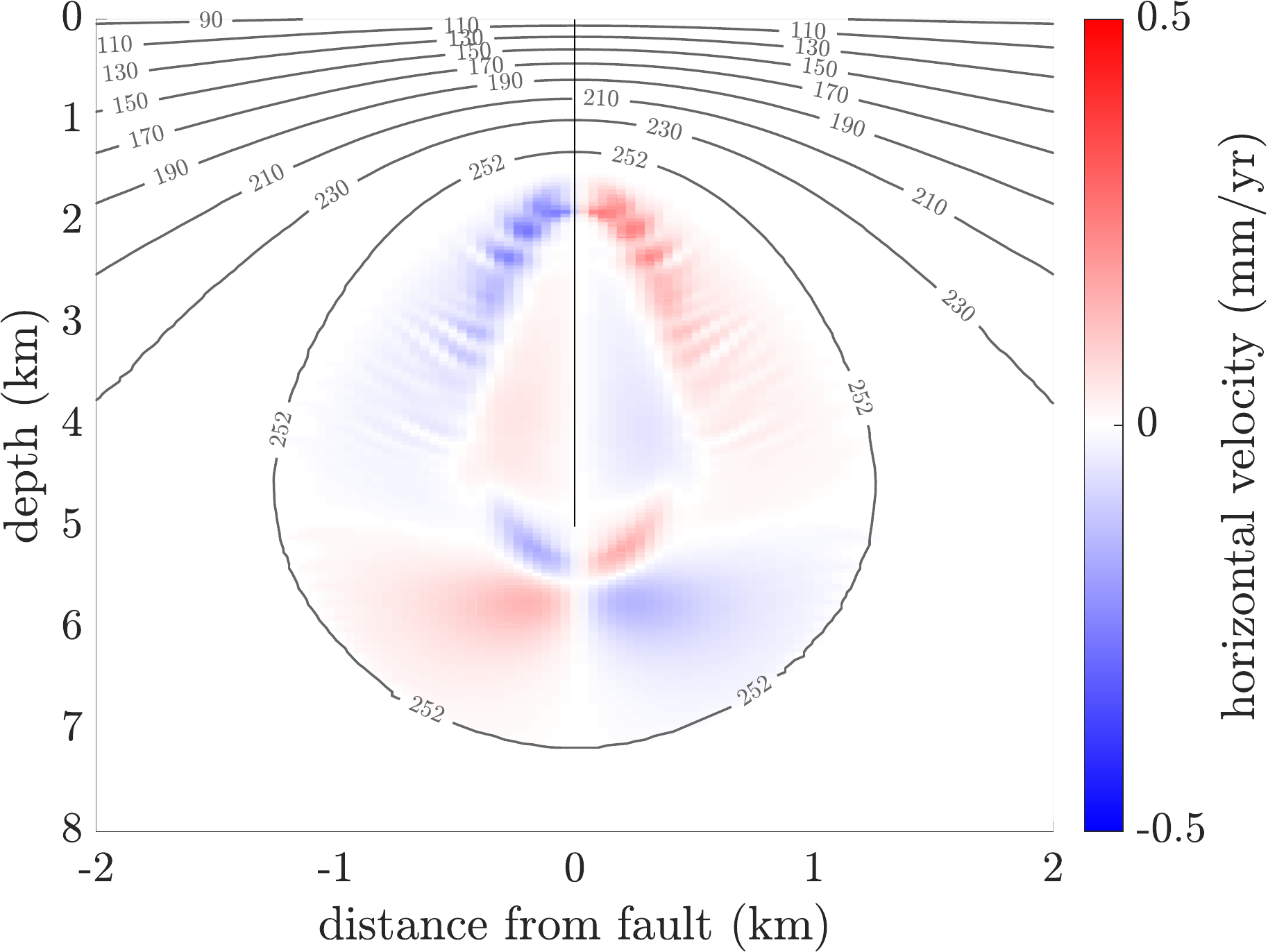}
    \caption{Flow velocity in the mushy zone for $F=500$: (left) vertical velocity, showing buoyant fluid rising along the crack and (right) horizontal velocity, demonstrating the buoyant plume is stable at $\mathtt{Ra_C}=200$ \citep{Bou2021} with periodic structures in the outer region of the mushy zone.}
    \label{fig:velocity}
\end{figure}

The fact that peak melt rates exceeds the observed geyser flux is an exciting demonstration of the viability of the shear heating model for the genesis of plumes on Enceladus and other icy satellites. While the simple configuration with steady heating doesn't account for the inherent complexities of the transient evolution, our impulse-response calculations likely elucidate the order of magnitude and timescale of heating. In figure \ref{fig:3tempandporosity}, we show the steady state temperature and porosity for three increasing nondimensional flux values, $F=180$, $F=240$, and $F=420$. These figures mirror figures 2(right) and 3(right) in the main text, where we show the temperature and porosity structure for $F=500$. For $F=260$, the maximum rate of liquid volume production is on the order of $0.05$ m$^{3}$ s$^{-1}$, which is less than rate required to indefinitely produce plumes with the observed ejection rate, but the mushy zone liquid reservoir could sustain geysers for tens of thousands of years. This suggests that there is a parallel to the Enceladus heat signal, which may be an indication of oscillations in thermal activity and hence plume activity \citep{Spe2013,Nim2018}. In other words, the liquid volume production rate by shear heating may allow for a large reservoir to develop periodically, leading to geysers that transiently exhausts the supply. Lastly, in our simulations, we reduce the shear heating rate as liquid is generated in the vicinity of the fracture. If we were to remove the liquid as geyser material, however, the shear heating could stay high, leading to additional liquid volume production. We currently do not remove the liquid from our simulation domain and leave these simulations to future work.

\begin{figure}[!ht]
    \centering
    \begin{overpic}[width=0.32\linewidth]{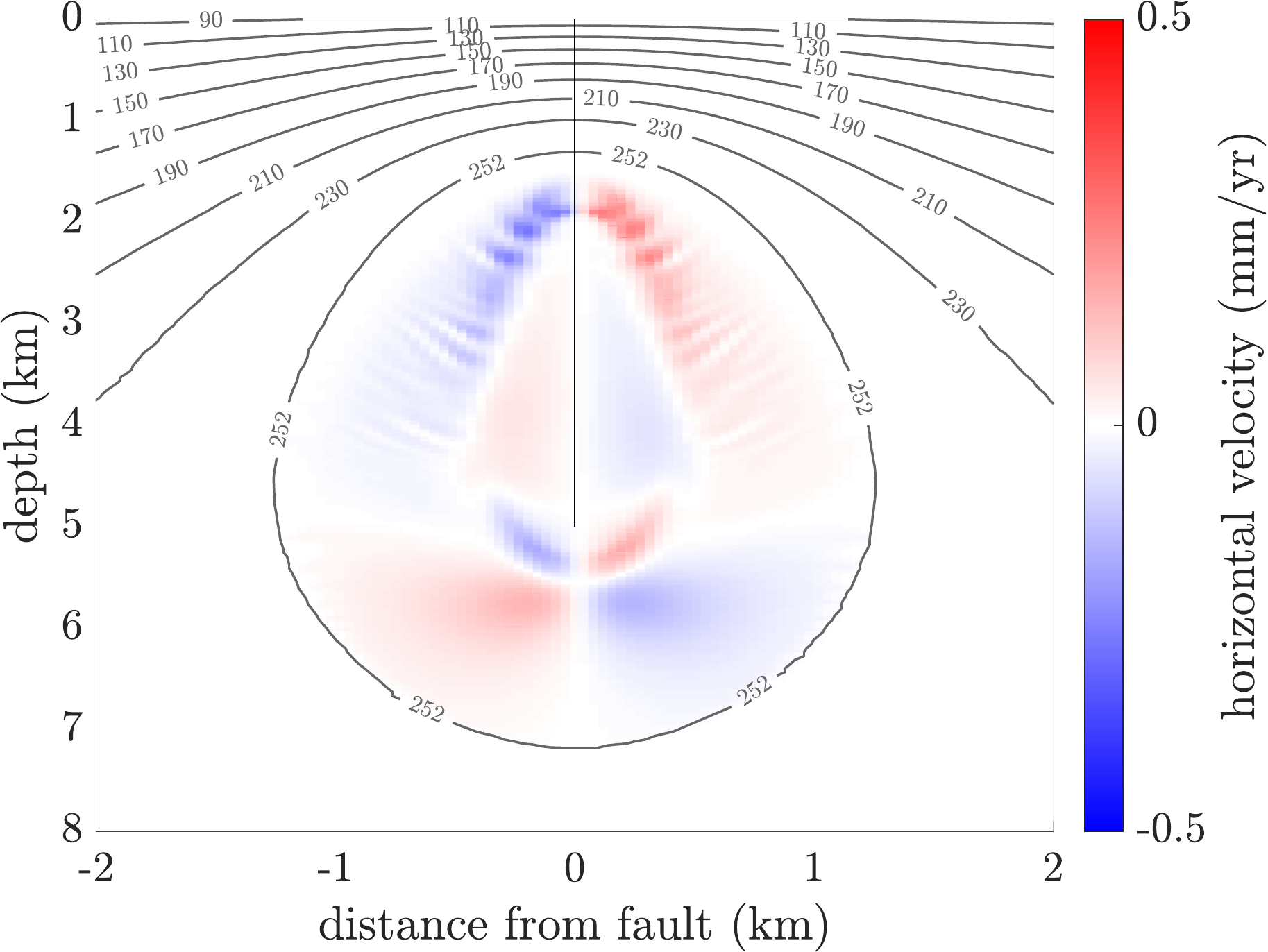}\put(33,76){$\mathtt{Ra_C} = 200$}\end{overpic}
    \begin{overpic}[width=0.32\linewidth]{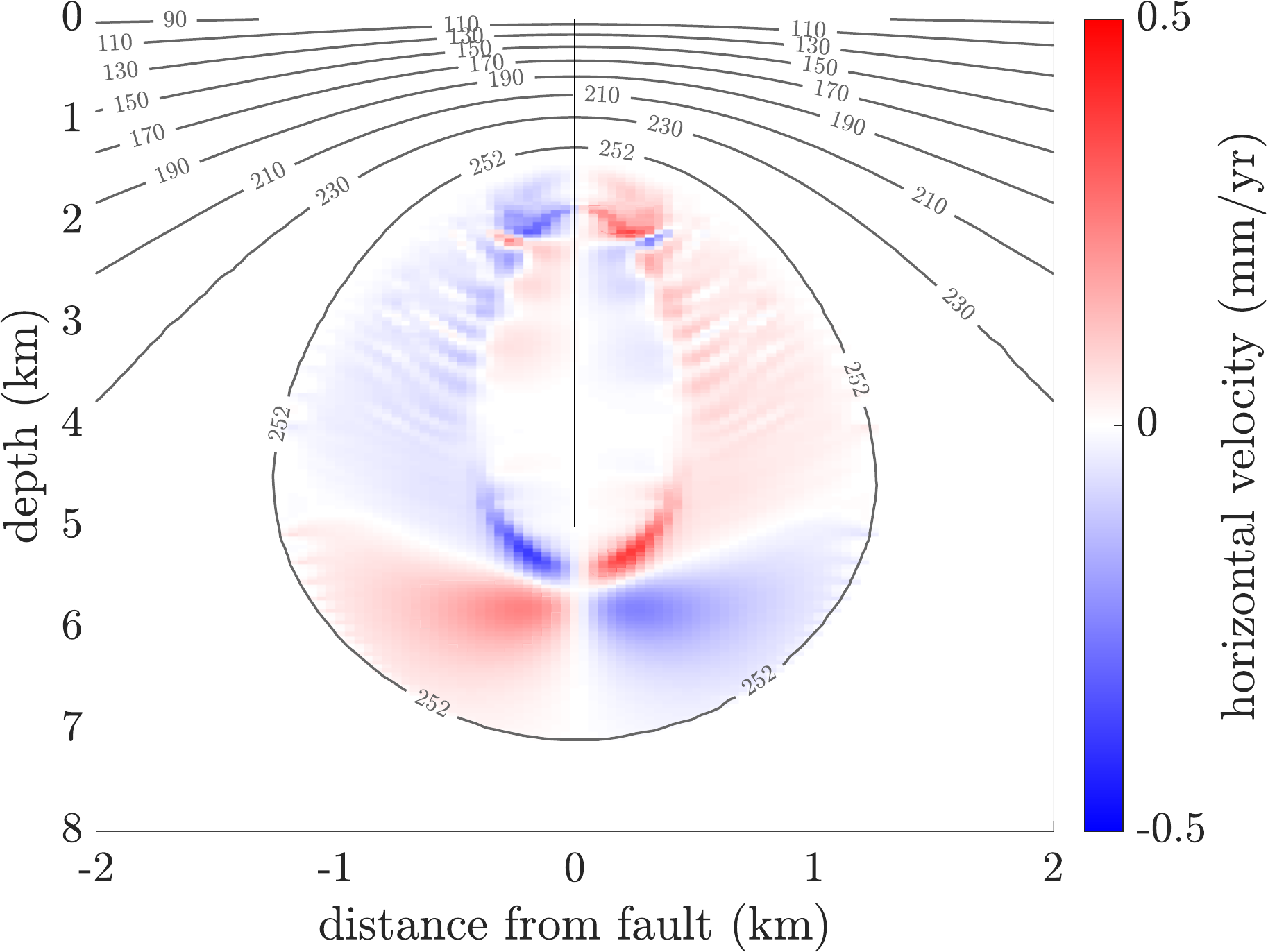}\put(33,76){$\mathtt{Ra_C} = 400$}\put(22,11){not converged}\end{overpic}
    \begin{overpic}[width=0.32\linewidth]{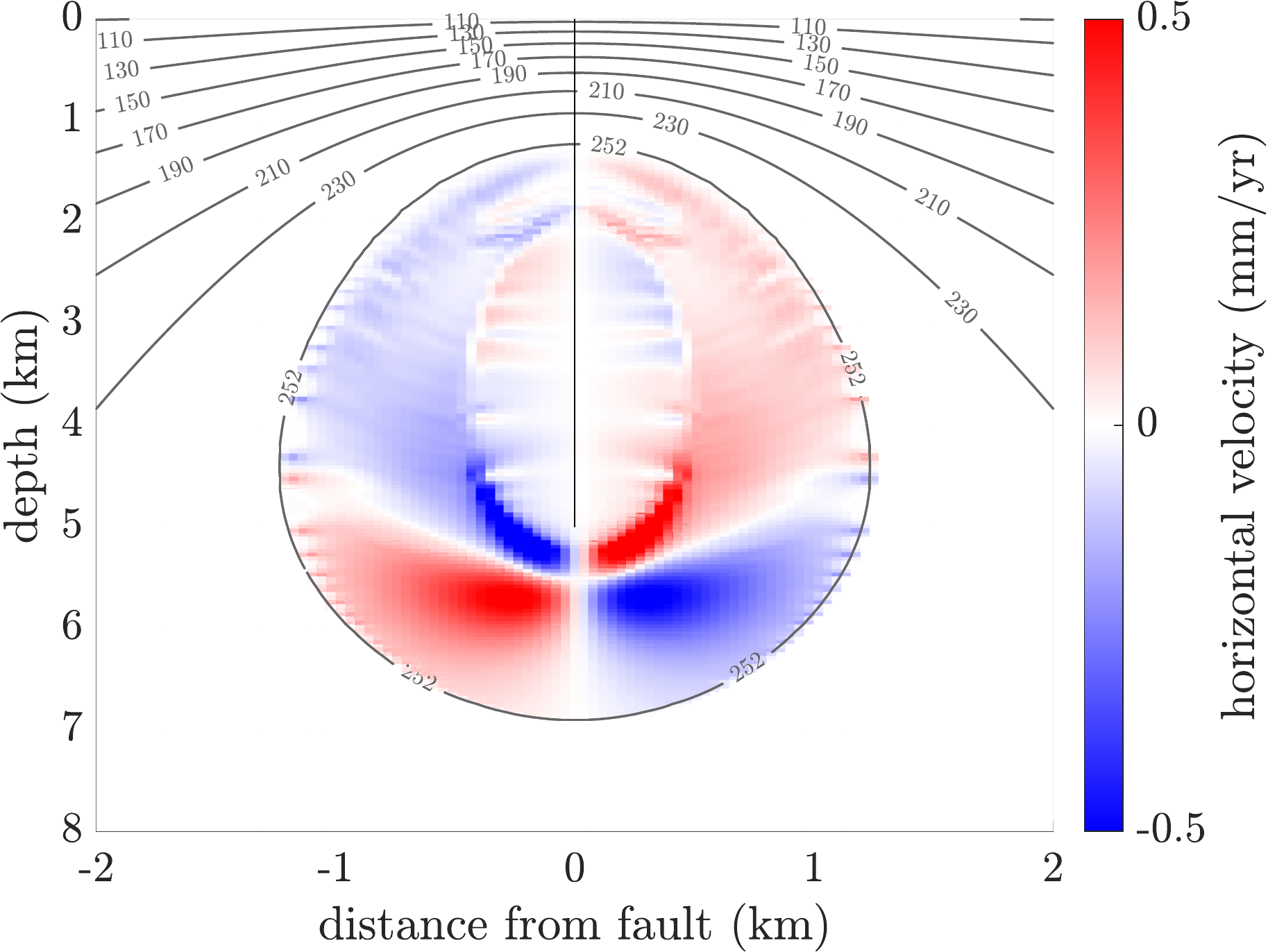}\put(33,76){$\mathtt{Ra_C} = 800$}\put(22,11){not converged}\end{overpic}
    \includegraphics[width=0.32\linewidth]{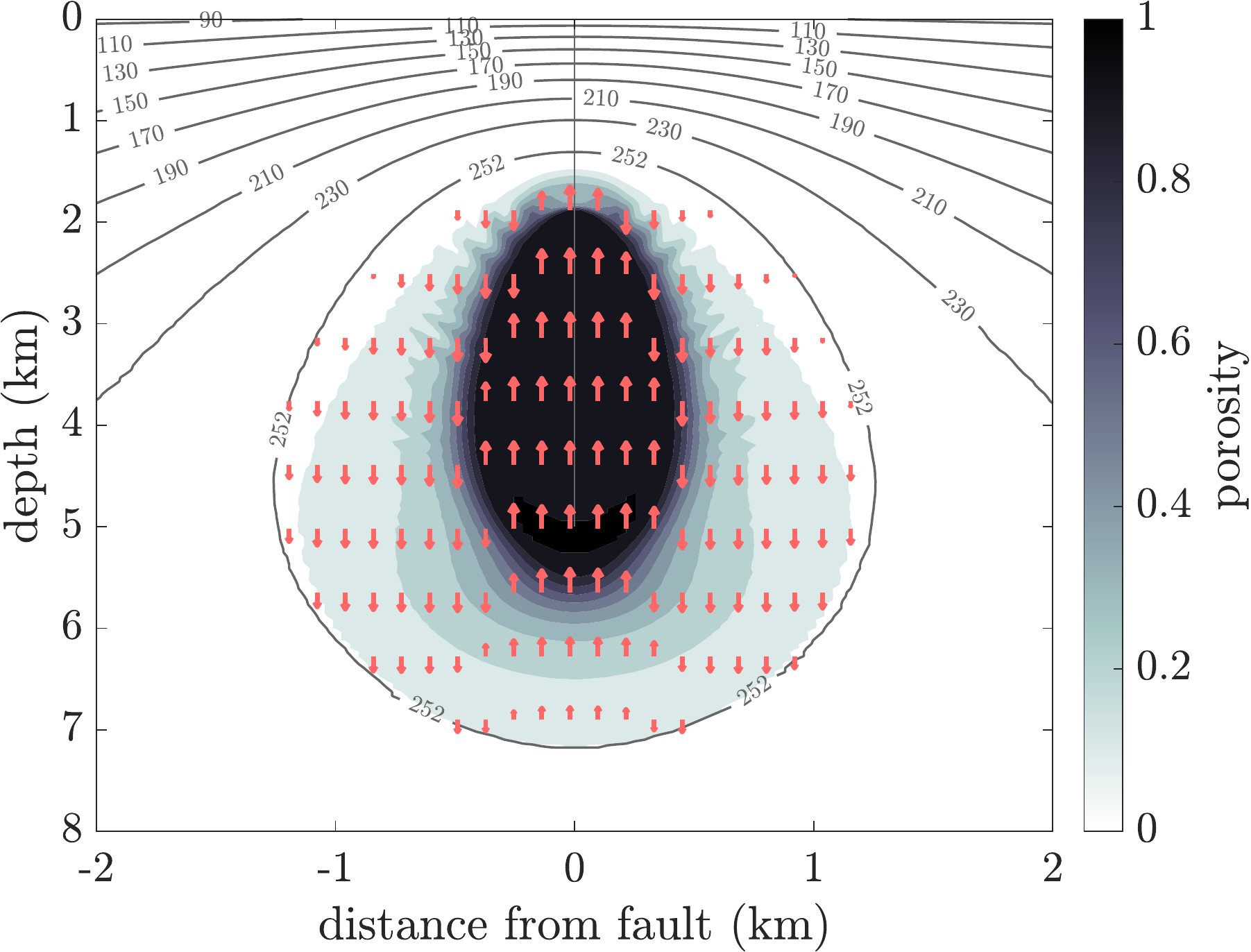}
    \includegraphics[width=0.32\linewidth]{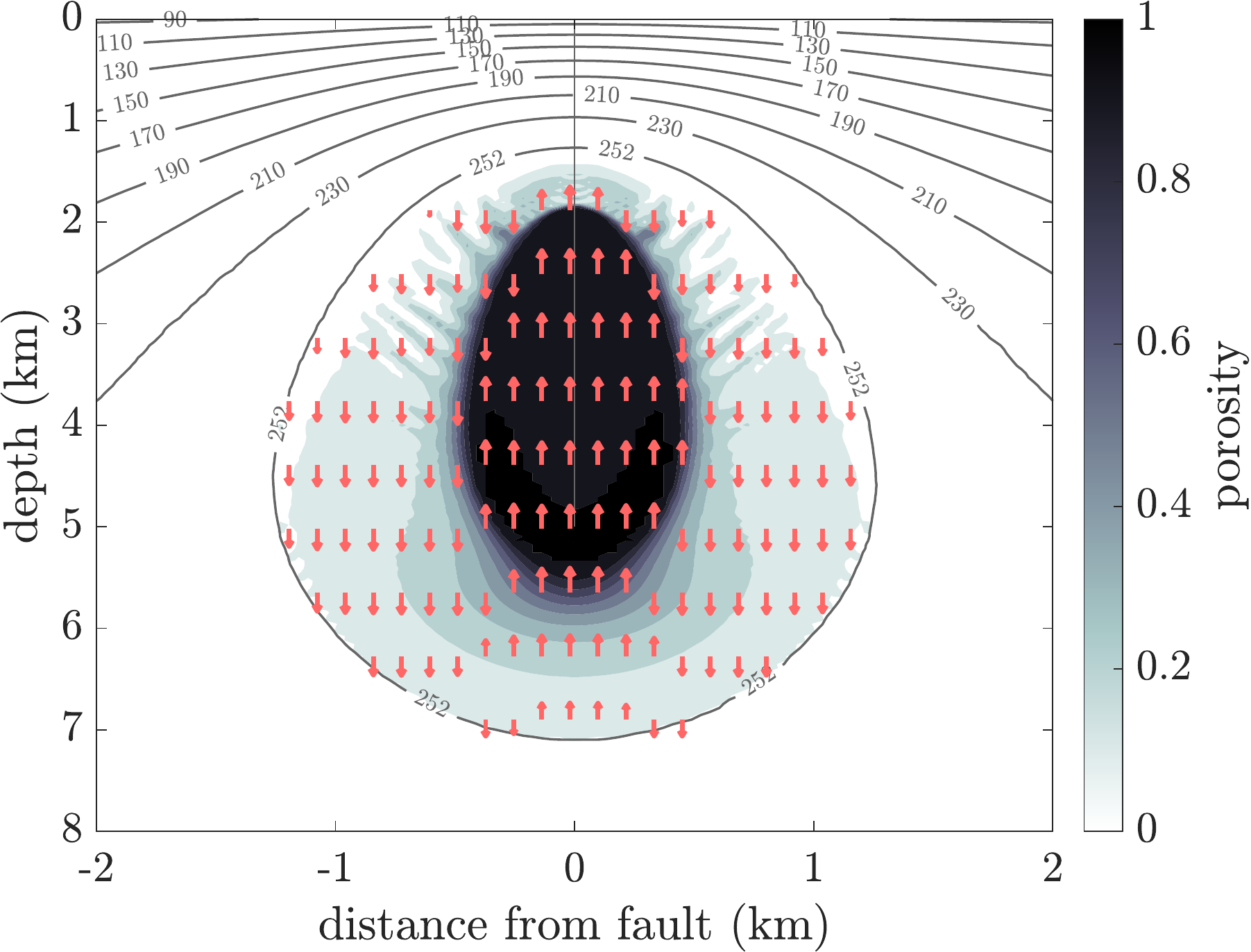}
    \includegraphics[width=0.32\linewidth]{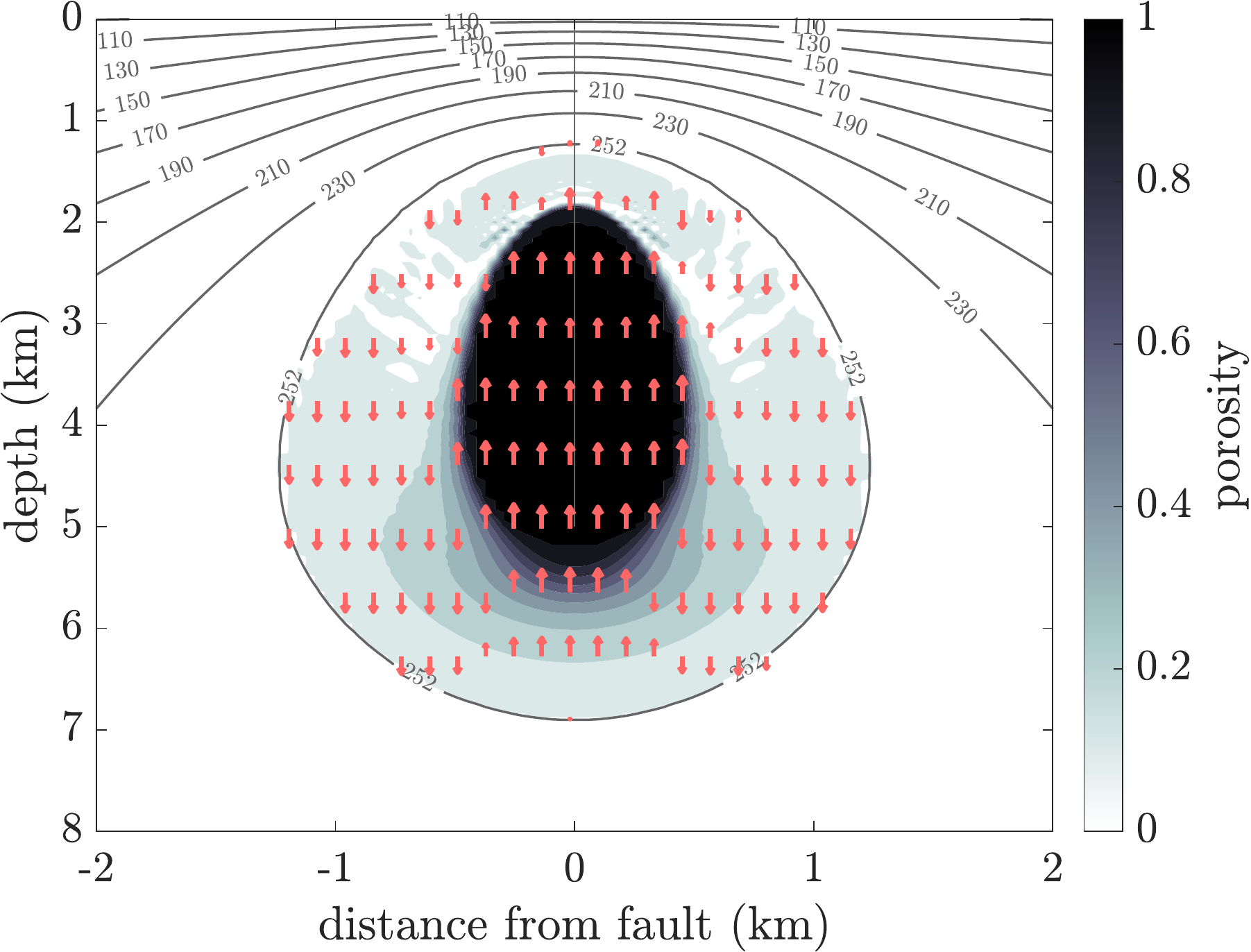}
    \caption{Increasing instability with stronger convection for $F=500$: (left) $\mathtt{Ra_C}=200$ showing a stable interior plume and an onset of instability in the region around the liquid inclusion. (middle) $\mathtt{Ra_C}=400$ showing an onset of instability in the interior plume. (right) $\mathtt{Ra_C}=800$ showing an unstable interior plume and a stable outer region that becomes unstable at the edge of the mushy zone. The three porosity plots, however, are similar.}
    \label{fig:RayleighNumber}
\end{figure}

Within the mushy zone, there is convection of the pore fluid and the fluid near the fracture rises buoyantly, as shown in the main text and in figures \ref{fig:velocity} and \ref{fig:RayleighNumber}. The intensity of convection is controlled by the thermal and compositional mushy Rayleigh numbers,
\begin{eqnarray}
    \mathtt{Ra_T} = \frac{\rho_{w} g \alpha \Delta T K_0 h}{\kappa_{o} \eta},\\
    \mathtt{Ra_C} = \frac{\rho_{w} g \beta \Delta C K_0 h }{\kappa_{o} \eta} ,
\end{eqnarray}
respectively. \citet{Bou2021} show that the rising plume along the fracture is unstable, with the development of the instability here resulting in rapid variation in horizontal velocity leading to numerical convergence issues in SOFTBALL (figure \ref{fig:RayleighNumber}). To find solutions that converge in a reasonable amount of time for our grid resolution, we follow \citet{Bou2021} and restrict the Rayleigh numbers to values below the instability threshold, $\mathtt{Ra_C}\approx205$ for our system and use $\mathtt{Ra_C}=200$ for the results we present. There is a large uncertainty in the permeability prefactor $K_0$ and the Rayleigh numbers that we present use a value of $K_0$ that is on the lower end for sea ice simulations \citep{Pol2017,Buf2018,Bou2021}. The low permeability limits convection and allows the simulations to converge numerically. In figure \ref{fig:RayleighNumber}, we compare three increasing Rayleigh numbers $\mathtt{Ra_C}=200,~400,~\mbox{and}~800$, where the latter two simulations do not converge. The higher Rayleigh numbers show that the flow-and-phase-change instability leads to fine-scale  structures that ultimately approach the grid resolution, affecting the pattern of the horizontal velocity field. The porosity field, however, is similar across the three Rayleigh numbers, suggesting that the results for total volume and peak melt rate would approximately hold as the Rayleigh number increased. 

\begin{figure}
    \centering
    \includegraphics[width=0.49\linewidth]{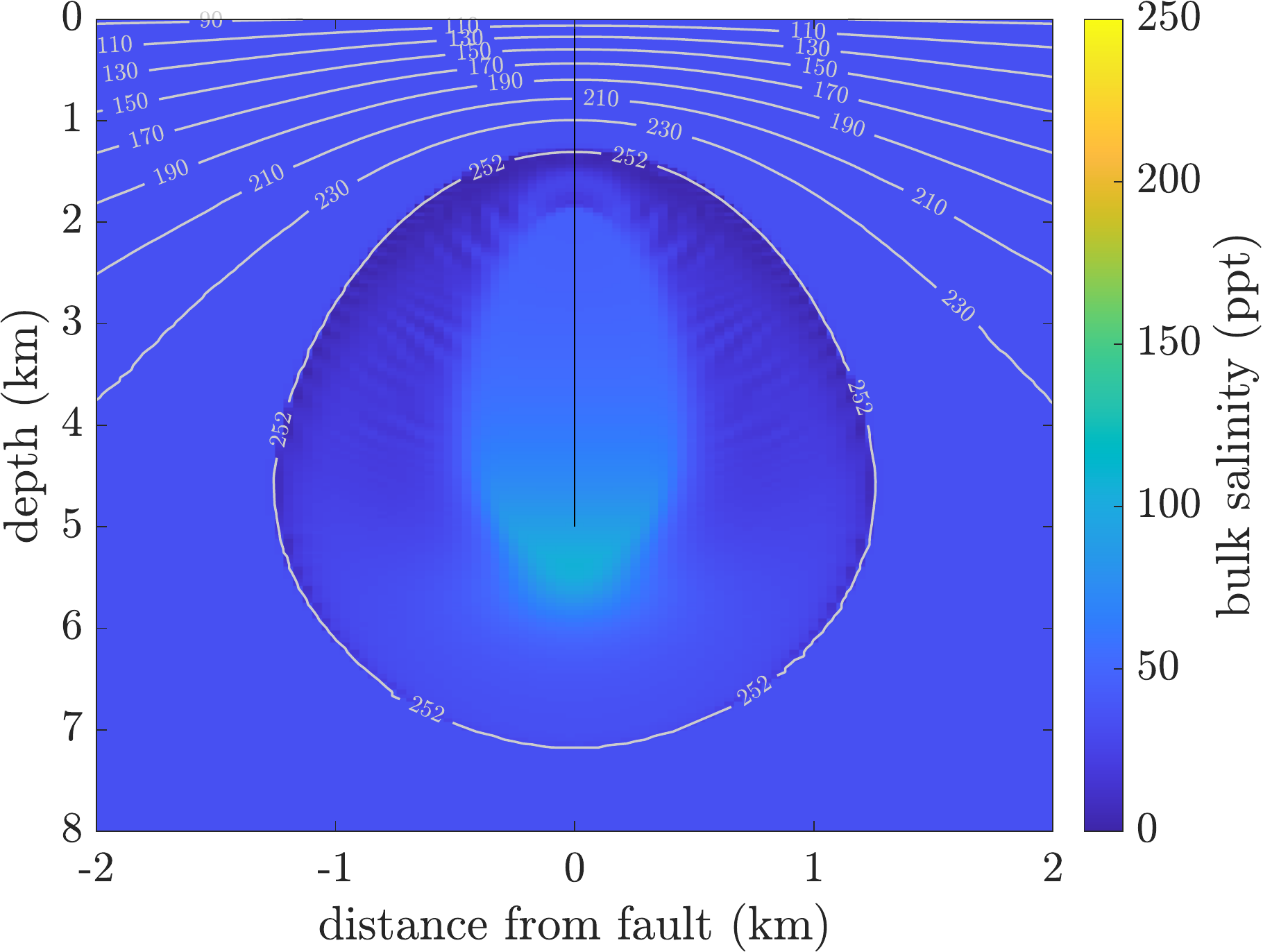}
    \includegraphics[width=0.49\linewidth]{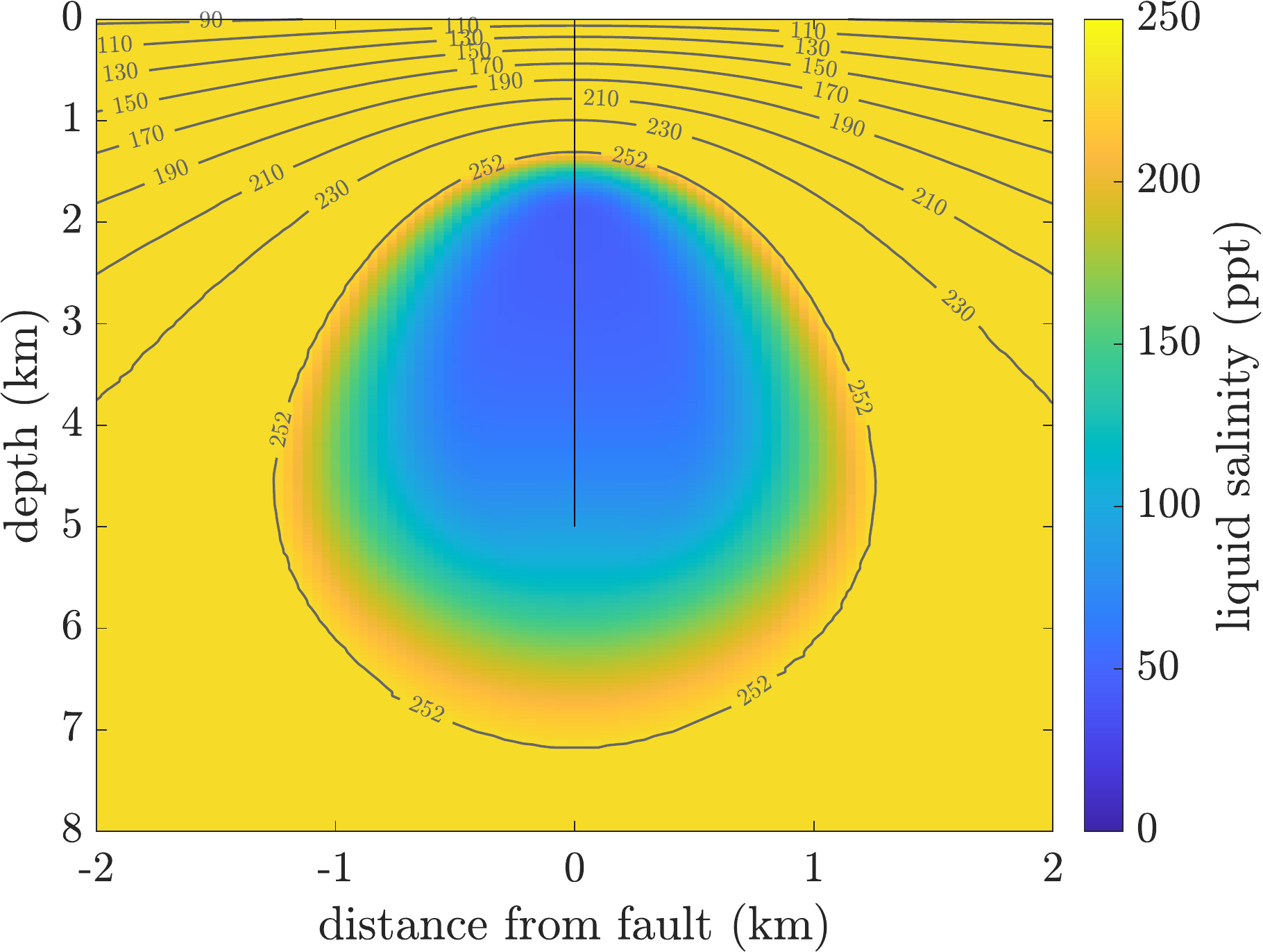}
    \caption{Salinity for the mushy zone with $F=500$. (left) Bulk salinity $C$, showing that the fluid within the interstices is saltier than the surrounding shell, which froze from the underlying ocean. (right) Liquid salinity $C_{\ell}$ showing the increase in salinity with depth in the mushy zone. High-salinity channels form at an oblique angle, due to the competition between melting and gravity drainage \citep{Buf2023}.}
    \label{fig:salinity}
\end{figure}

\begin{figure}
    \centering
    \includegraphics[width=0.324\linewidth]{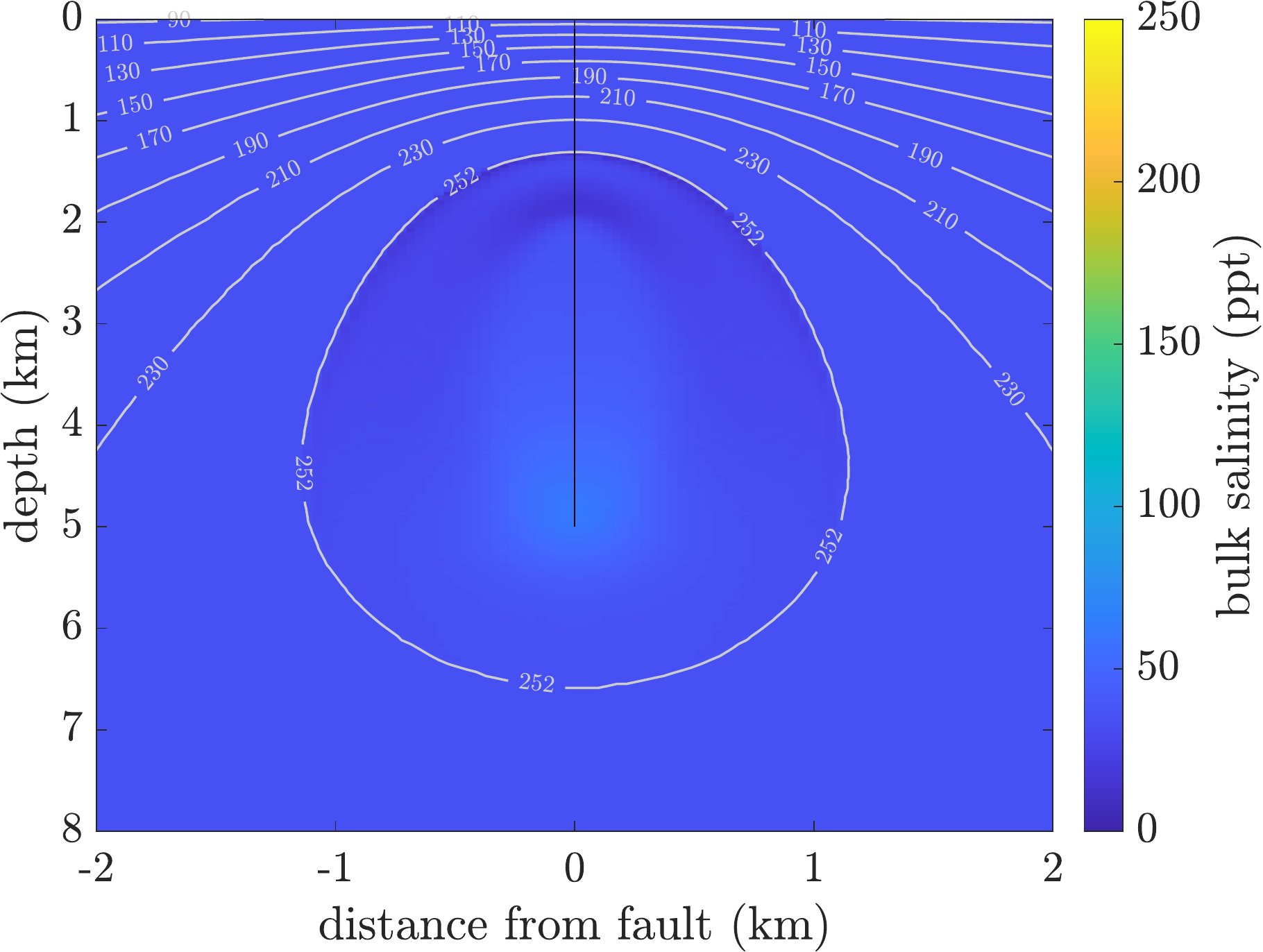}
    \includegraphics[width=0.324\linewidth]{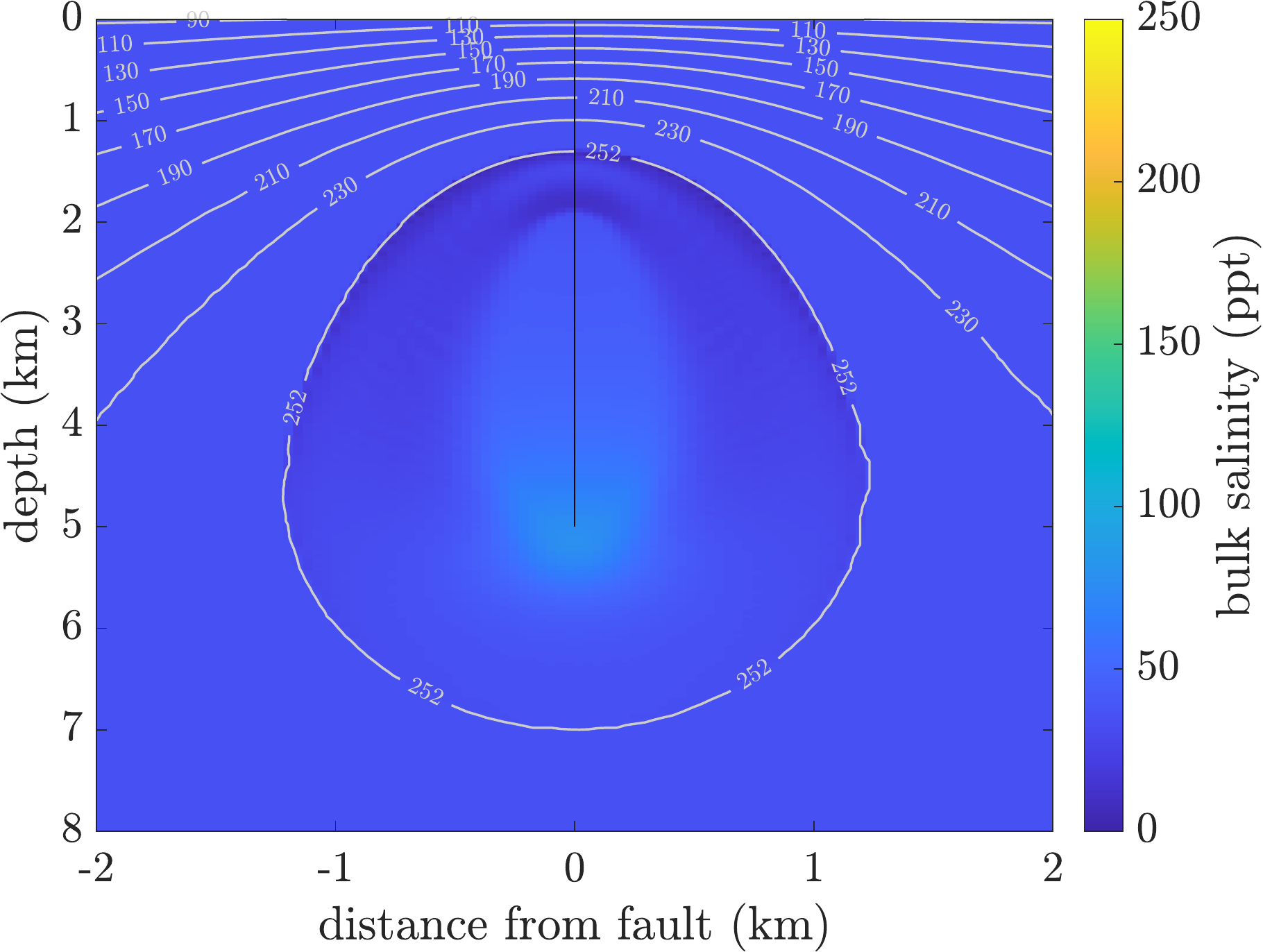}
    \includegraphics[width=0.324\linewidth]{pltU_salinity500.pdf}
    \caption{Bulk salinity evolution with $F=500$, showing the drainage and brine concentration along the fracture: (left) 541 kyr. (middle) 865 kyr. (right) 13,000 kyr.}
    \label{fig:salinityevolution}
\end{figure}

Near the fracture, shear heating melts the ice and dissolves the salt. Therefore, the pore fluid in this region has a higher salinity (figure \ref{fig:salinity}) than the surrounding shell. Here we plot the bulk salinity $C$, which is given as
\begin{equation}
    C = \phi C_\ell + (1-\phi)C_s,
\end{equation}
where $\phi$ is the porosity, $C_{\ell}$ is the liquid salinity, and $C_s$ is the solid salinity. Starting from a step change in heating, the evolution of the bulk salinity is shown in figure \ref{fig:salinityevolution}. First, shear heating warms the salty ice above the eutectic temperature generating a permeable zone. Since the brine salinity and temperature are connected through the liquidus constraint within the mushy zone, colder and saltier brine along the top then sinks and the resulting convective desalination reduces the bulk salinity when compared to the background value. Next, the sinking brine cannot escape and collects near the base of the porous inclusion along the fracture, generating a salt-rich zone with porosity near unity. Convection continues within the mushy zone. The convective desalination leaves behind regions of high bulk salinity in the outer parts of the porous inclusion, that are trapped due to the reduction of permeability in the neighbouring desalinated ice which now has low porosity. Due to the partial melting within the mushy zone, the chemical composition of the geyser ejecta sourced from the shear heating model that we propose here will be different from geyser ejecta sourced from the ocean. In our simulations, the salinity in the fluid around the crack is higher than the background shell salinity. The shell likely froze from the underlying ocean and will have a salinity structure that varies with depth \citep{Buf2021b}.

\section{Dike propagation\label{sec:dike}}
In our steady state simulations, a region of salty brine develops along the fracture with a porosity approaching a value of $\phi \approx 1$. Here we do not allow the liquid to escape as geysers, therefore the liquid volume can build up within the mushy zone. The question then arises as to whether the brine pocket can hydraulically fracture through the ice shell and into the underlying ocean, thereby generating a dike that opens a surface-to-base connection. We analyze this possibility through a scaling analysis, following \citet{Lis1990} and \citet{Kal2016}. For $F=500$, we take the half width of brine region to be $a\sim200$ m, the breadth into the page as $w\sim 500$ km, and the height as $\ell\sim4$ km, we have two options for the balance of gravity. Either, viscous stresses resist propagation or the fracture toughness $K_I\sim 100$ kPa m$^{1/2}$ resists opening. Starting with the case of viscous stress, we have that 
\begin{equation}
    \phi \Delta \rho g a \ell w \sim \mu \frac{V}{a^2} a \ell w, 
\end{equation}
where $\phi\sim0.25$ is a representative average porosity in the brine pocket; $\Delta \rho \sim 90$ kg m$^{-3}$ is the density difference between the brine and the surrounding ice; $\mu\sim10^{14}$ Pa$\cdot$s is the brine-saturated temperate ice viscosity; and $V$ is the rate of dike propagation. Canceling terms, we find that the rate of propagation scales as
\begin{equation}
    V \sim \frac{\phi \Delta \rho g a^2}{\mu}.
\end{equation}
The time $t_{\textrm{dike}}$ for a dike to propagate a distance $\ell$, then scales as
\begin{equation}
    t_{\textrm{dike}} \sim \frac{\ell}{V} \sim \frac{\mu \ell}{\phi \Delta \rho g a^2} \sim 110~\mbox{kyr}.
\end{equation}
This timescale is on the same order of magnitude as the time to steady state in our shear heating calculations, indicating that the brine pocket could form and then propagate as a dike within that time frame, if the porosity grows to order 1 without significant venting of the liquid from the crack. 

If the balance is instead between gravity and the fracture toughness, we have that 
\begin{equation}
    \phi \Delta \rho g a \ell w \sim \frac{K_I}{\ell^{1/2}} a w, 
\end{equation}
which determines whether the fracture will propagate or not. Simplifying and plugging in numbers, we have that
\begin{equation}
    \phi \Delta \rho g \ell^{3/2} \sim 700~\mbox{kPa m$^{1/2}$}~~>~~K_I\sim 100 ~\mbox{kPa m$^{1/2}$}. 
\end{equation}
Thus, we find that $\phi \Delta \rho g \ell^{3/2}$ is greater than the ice fracture toughness $K_I$ and much larger than values measured in brine-saturated sea ice, e.g. $\sim50$ kPa m$^{1/2}$ \citep{Sch2009}, demonstrating that the ice will not provide significant resistance to fracture.

In both cases, we see that dike propagation is possible if the porosity reaches order 1 over an extended region, before the liquid is vented. The time scale for diking is on the same order of magnitude as the for our simulations to reach a steady state and the ice fracture toughness is not large enough to prevent fracture. These scaling calculations spawn a few ideas. An internal dike is interesting because there could be periodic brine pocket generation, leading to opening a connection between surface and ocean, then \citet{Kit2016}-style eruptions. Or the dike could drain all of the fluid internally, leading to a stronger fault with larger shear heating, and a periodic mushy zone source for the geysers. However, by not allowing for interstitial liquid to escape, our estimates for the liquid brine volume are artificially high, meaning that a region with porosity approaching 1 may never form within the mushy zone. In this case, there would be no prospect for a dike and the material generated would only emanate out as geysers. We leave additional analysis exploring these ideas to future work. 

\begin{table}
\centering
\begin{tabular}{llllllll}
\hline
parameters\\
\hline
specific heat of ice & $c_i$ & 2000 J kg$^{-1}$ K$^{-1}$ \\
specific heat of ocean & $c_o$ & 4000 J kg$^{-1}$ K$^{-1}$\\
ice density & $\rho_i$ &  940 kg m$^{-3}$\\
seawater density & $\rho_w$ &  1030 kg m$^{-3}$\\
density coefficient for temperature & $\beta_T$ &  2.1$\times10^{-4}$ kg m$^{-3}$ K$^{-1}$\\
density coefficient for salt & $\beta_C$ &  7.7$\times10^{-4}$ kg m$^{-3}$ ppt$^{-1}$\\
thermal conductivity of ice & $k_i$ &  2.0 W m$^{-1}$ K$^{-1}$\\
thermal conductivity of ocean & $k_o$ &  0.6 W m$^{-1}$ K$^{-1}$\\
latent heat of fusion & $\mathscr{L}$ &  335000 J kg$^{-1}$\\
water viscosity & $\eta$ &  2$\times10^{-3}$ Pa s\\
salt diffusivity & $\kappa_s$ &  2$\times10^{-9}$ m$^2$ s$^{-1}$\\
thermal diffusivity & $\kappa_o$ &  1.5$\times10^{-7}$ m$^2$ s$^{-1}$\\
Stefan-Boltzmann constant & $\sigma$ & $5.67\times10^{-8}$ W m$^{-2}$ K$^{-4}$\\
\hline
shell salt composition & $C_i$ & 35 ppt \\
eutectic composition & $C_e$ & 230 ppt\\
liquidus slope & $m$ & -0.0913 K~ppt$^{-1}$\\
eutectic temperature & $T_e$ & 252 K\\
surface temperature & $T_s$ & 75 K\\
gravity on Enceladus & $g$ & 0.113 m~s$^{-2}$\\
fracture depth & $h$ & 5 km\\
height scale & $H$ & 10 km\\
permeability prefactor & $K_0$ & 3.2$\times10^{-13}$ m$^2$\\
Hele-Shaw cell spacing & $d$ & 5$\times10^{-5}$ m\\
time scale & $t_c$ & 2.2$\times10^7$ years\\
slip velocity & $u$ & 3.4$\times$10$^{-6}$ m s$^{-1}$\\
temperature scale & $\Delta T$ & 18 K\\
\hline
emissivity & $\epsilon$ & 1\\
coefficient of friction & $\mu$ & 0.3\\
Stefan number & $\mathtt{St}$ & 4.7\\
Prandtl number & $\mathtt{Pr}$ & 12\\
Lewis number & $\mathtt{Le}$ & 73\\
mushy compositional Rayleigh number & $\mathtt{Ra_C}$ & 200\\
mushy thermal Rayleigh number & $\mathtt{Ra_T}$ & 2\\
reluctance & $\mathtt{R}$ & 9.6\\
shear heating & $F$ & 500\\
linearized radiation, nondimensional $r_1$ & $G$ & 0\\
linearized radiation, nondimensional $r_2$ & $b$ & 100\\
linearized radiation, nondimensional $T_s$ & $\theta_s$ & -10\\
\hline
\end{tabular}
\caption{Parameters used in the SOFTBALL numerical simulations, grouped by constants, variables, nondimensional parameters. Definitions follow \citet{Par2020}. Values from \citet{Nim2002}, \citet{Nim2007}, \citet{Spe2013} as well as \citet{Buf2021b}.}
\label{prms}
\end{table}

\bibliographystyle{plainnat}
\bibliography{Glib}